\PassOptionsToPackage{dvipsnames}{xcolor}
\documentclass[twocolumn]{aastex631}
\usepackage[utf8]{inputenc}
\usepackage[T1]{fontenc}
\usepackage{amsmath,amssymb}
\usepackage{graphicx}
\usepackage{xcolor}
\usepackage{ulem}
\usepackage{float}
\hypersetup{colorlinks=true,allcolors=blue}

\graphicspath{{figures/}}

\renewcommand{\vec}[1]{\text{\mathversion{bold}$#1$}}
\newcommand{\diff}{\mathrm{d}}
\newcommand{\uv}[1]{\text{\mathversion{bold}$\widehat{#1}$}}

\begin{document}

\title{Gravitational-Wave Signal from a {Short} GRB Relativistic MHD Jet}

\author{Or Zuriel}
\affiliation{Racah Institute of Physics, The Hebrew University of Jerusalem, Jerusalem 91904, Israel}

\author{Dmitry D. Ofengeim}
\affiliation{The Raymond and Beverly Sackler School of Physics and Astronomy, Tel Aviv University, Tel Aviv 69978, Israel}
\author{Omer Bromberg}
\affiliation{The Raymond and Beverly Sackler School of Physics and Astronomy, Tel Aviv University, Tel Aviv 69978, Israel}
\author{Eyal Efrima}
\affiliation{Racah Institute of Physics, The Hebrew University of Jerusalem, Jerusalem 91904, Israel}
\author{Tsvi Piran}
\affiliation{Racah Institute of Physics, The Hebrew University of Jerusalem, Jerusalem 91904, Israel}

\begin{abstract}
We analyze the gravitational-wave (GW) signal produced by a relativistic magnetohydrodynamic (MHD) simulation of a short GRB jet with parameters corresponding to those of GRB 170817A. The signal that is generated in the hidden central engine of the GRBs is computed by direct integration of the retarded solution of the linearized Einstein equations of an MHD simulation. Separating the GW signal into its baryonic and electromagnetic contributions, we identify the physical origin of its dominant features. The waveform is of the memory type, with $|\tilde h(f)| \propto 1/f$ at low frequencies and a knee at $\sim 0.3$–$0.5$ Hz set by the energy injection time, which corresponds in this case to the GRBs duration. A pronounced early peak reflects the opening of the magnetic field lines and the equatorial kick given to the ejecta at the onset of the jet, demonstrating that the GW signal can reveal processes deep in the GRB engine that cannot be observed otherwise. The signal is strongest for observers far from the jet axis. The total radiated GW energy is only $\sim 10^{42}$ erg, about $10^{-9}$ of the jet energy; it scales as ${\cal E}^2/\tau$ with the shortest variability time $\tau$, so this is a lower limit. Comparing the predicted spectrum with the sensitivity of current and future detectors, we find that events similar to GW170817 are marginally detectable by DECIGO from up to 10 Mpc and by the Einstein Telescope from 1 Mpc. The injected Poynting flux in this simulation was smooth. An engine variable on a millisecond time scale, as predicted in some GRB models, would increase the GW energy and detectability substantially. 

\end{abstract}


\section{Introduction}
\label{sec:intro}

The direct detection of gravitational waves has opened a new observational
window on energetic astrophysical phenomena. All currently detected GW events
originated from  compact-object mergers and the signals observed so far are dominated by the in-spiral phase with some contribution from  the coalescence moment and the subsequent ringdown.  Binary neutron star mergers also produce short GRBs involving relativistic jets \citep{Eichler1989}. These jets  are also expected to
generate gravitational radiation through their highly anisotropic, time-dependent stress-energy distribution. Jet-generated gravitational waves
are particularly interesting because they provide direct information that cannot be revealed otherwise about the jet acceleration mechanism and the central engine.

About {twenty-five} years ago   \citet{segalis2001} and
\citet{piran2002} pointed out that the acceleration of a Gamma-ray burst (GRB) jet to relativistic velocities produces gravitational waves. These authors
considered a point
mass that is instantaneously accelerated to a Lorentz factor $\Gamma$, with
total energy ${\cal {E}} = m\Gamma c^2$. In this zero-frequency limit the waveform is
a step function, i.e.\ a \emph{memory-type} signal  with an  amplitude   $ h(\theta_v) = {2 G {\cal E}  \beta^{2}}/{(c^{4} d)} \cdot 
[{\sin^{2}\theta_v}/({1-\beta\cos\theta_v})]$, where $d$  is the distance to the source, $\theta_v$ is the observer's viewing angle relative to the jet axis and $G$ and $c$ are Newton's constant and the speed of light. These studies also demonstrated another important aspect of this signal - {\it antibeaming}. 
The signal  vanishes along the direction of
motion, reaches half of its maximum at $\theta_v = \Gamma^{-1}$ and peaks at
$\theta_v \simeq \sqrt{2/\Gamma}$.
The instantaneous point-particle picture gives the final jump of the amplitude but none of the temporal structure that determines the unique characteristics of the signal.  

\citet{sago2004}  described a GRB as a
large number of thin instantly accelerated ``minijets'' that are ejected at random angles within a
cone and at random times during the burst. 
The superposition of these minijets {spreads} the signal so that instead of a Heaviside function  it rose on a time scale comparable to the duration of the GRB. \citet{birnholtz2013} relaxed
the assumption of instantaneous acceleration and calculated the signal of a
continuously accelerating axisymmetric jet. Integrating over
equal-arrival-time surfaces for different angular structures and viewing
angles, they found again a temporal structure of the order of the acceleration time for observers close to the jet axis, and longer at larger angles. \citet{leiderschneider2021} combined prolonged acceleration and
continuous ejection, and identified the properties of the signal that do not
depend on a particular acceleration model. The waveform is governed by two
time scales, the acceleration time $t_{acc}$ and the injection time $t_{inj}$,
and its spectrum changes at a crossover frequency $f_c = 1/\max\left[t_{inj},\, (1-\beta\cos(\theta_v+\theta_j))\,t_{acc}\right]$
where $\theta_j$ is the jet opening angle. The Fourier transform is $\propto 1/f $ below $f_c$.  Above this crossover frequency it decreases faster reflecting the details of the source and viewing angle (see \citealt{piran2022} for a review).

All these studies prescribe schematic descriptions for the jet rather than consider realistic configurations.  
They consider only a matter outflow, whereas the jets could be Poynting flux dominated. Most importantly these initial works ignore the essential interaction of the jet with its surroundings that shapes its overall shape and final Lorentz factor. 
\cite{Sakai2025} considered memory type GW waves from interacting shell, following the internal shocks model. 
{GW emission from realistic models of long GRBs has recently been addressed by, e.g., \cite{Urrutia+MNRAS2023,Urrutia+2026_Hidden}.}
In the following we present such a simulation of a short GRB jet under conditions that mimic the parameters of the short GRB 170817A.

The calculations of the GW signal are complicated by the fact that the jet is relativistic. Hence the common quadrupole approximation is not valid. A full general relativistic simulation is typically too demanding but at the same time it is not needed as the general relativistic effects involved in the jet simulations are minimal and the back reaction of the emitted GW on the jet is negligible.
A complication is, however, to account for a relativistic effect of time retardation.
To resolve this we developed \citep{zuriel2026} a dedicated GPU code that follows a jet simulation and calculates the retarded solution of the linearized Einstein equations from the stress energy momentum of the source.

The structure of the paper is as follows. We begin in \S~\ref{sec:analytic} by a brief description of the GW equation, pointing out an important feature of the memory type signal that enables us to distinguish between different sources of the obtained GW signal. We continue in \S~\ref{sec:methods} with a short description of the numerical {scheme} and follow in \S~\ref{sec:test} with code tests, that also shed some light on the behaviour of jet-GW. We present  the jet simulations in \S~\ref{sec:jet} and our  results 
in \S~\ref{sec:results}. We summarize the results and consider the detectability  of such signals in \S~\ref{sec:conclusions}. 
We use $G=c=1$. Metric signature is $-,+,+,+$. For a vector $\vec{a}$, a co-directed unit vector is $\uv{a}$. Greek indices for $0,1,2,3$, Latin indices for $1,2,3$.

\section{Analytic considerations}
\label{sec:analytic}
Because of the relativistic motion within the source the gravitational-wave signal computed using the direct 
integration of the retarded solution of the linearized Einstein equations in the harmonic gauge yields the signal measured by an observer at $\vec{r}_\text{obs}$ at the moment $t'$ \citep{WeinbergBook1972}
\begin{equation}
\begin{aligned}
    h_{\mu\nu}(\vec{r}_\text{obs},{t'}) &= 4\int \diff^3\vec{r}\, \frac{S_{\mu\nu}(\vec{r}, t_r)}{|\vec{r}_\text{obs} - \vec{r}|},
    \\
    S_{\mu\nu} &= T_{\mu\nu} - \frac{1}{2}\eta_{\mu\nu} T_\lambda^\lambda,
    \quad
    t_r = t' - |\vec{r}_\text{obs} - \vec{r}|.
\end{aligned}
    \label{eq:LienardWiechert}
\end{equation}
Here $h_{\mu\nu}$ is the gravitational perturbation of the metric tensor, $T_{\mu\nu}$ is the energy-momentum tensor of the source, and $t_r$ is called the retarded time. For astrophysical sources, the far zone approximation, $|\vec{r}_\text{obs}| = D \gg r$ works perfectly. Then one can neglect $\vec{r}$ in the denominator, and for the retarded time it is enough to write $t_r \approx {t'} - D + \uv{n}\cdot\vec{r}$, where the last term, with $\uv{n} = \vec{r}_\text{obs}/D$, takes the only relevant dependence on the spatial distribution of the source. {Hereafter we drop the constant propagation delay $D$ from the observer time, so that $t_r = t' + \uv{n}\cdot\vec{r}$, and $t'=0$ corresponds to the arrival of a signal emitted from the origin at $t=0$. Throughout the paper $t$ denotes the laboratory time of the source and $t'$ the observer time.}

\begin{figure}
    \centering
    \includegraphics[width=1.0\linewidth]{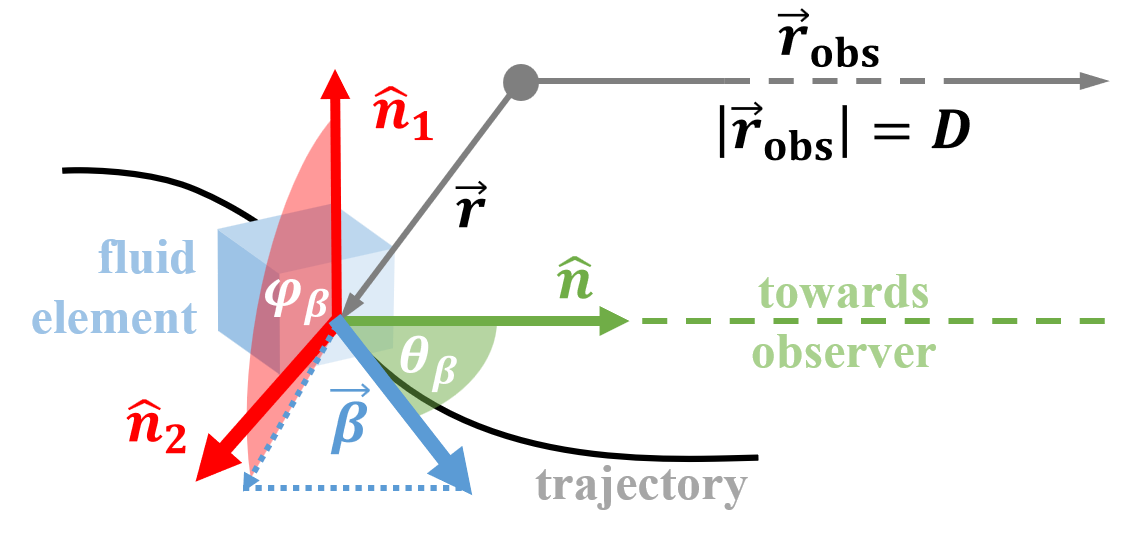}
    \caption{Illustration of the basic terms to describe emission and detection of GWs. The source is assumed to be a material object with velocity $\vec{\beta}$.}
    \label{fig:Vector_basis}
\end{figure}

For the GW emission, there are only two physical degrees of freedom in the tensor $h_{\mu\nu}$, which are given by the plus and cross polarization mode components. Hereafter we consider only the first one \citep{MTW2017}:
\begin{equation}
    \label{eq:h+fromT}
    h_+ = \frac{2}{D} \int \diff^3\vec{r} \left( n_{1i} n_{1j} - n_{2i} n_{2j} \right) T_{ij}(\vec{r},t_r).
\end{equation}
Unit vectors $\uv{n}_1 \perp \uv{n}$ and $\uv{n}_2 = \uv{n}\times\uv{n}_1$ define the detector orientation and do not depend on properties of the GW source. Together with the direction to the observer $\uv{n}$, these vectors form a right triad (see Fig.~\ref{fig:Vector_basis}). 

For an astrophysical source that consists of a magnetized plasma, the energy-momentum tensor has two contributions, material fluid  and electromagnetic , with the spatial components
\begin{equation}
    \label{eq:Tij}
\begin{aligned}
    T_{ij}^{(matter)} &= (\varepsilon + P) \Gamma^2 \beta_i \beta_j + \delta_{ij} P,
    \\
    T_{ij}^{(EM)} &= - \frac{E_i E_j + B_i B_j}{4\pi} + \delta_{ij} \frac{E^2 + B^2}{8\pi}.
\end{aligned}
\end{equation}
Energy density $\varepsilon$ and pressure $P$ are defined in the rest frame of the fluid, $\vec{\beta}$ is its 3-dimensional velocity, and $\Gamma = 1/\sqrt{1-\beta^2}$. The electric field $\vec{E}$ and magnetic field $\vec{B}$ are defined in the laboratory frame. Then the total GW signal from the relativistic astrophysical source is $h_+ = h_+^{(matter)} + h_+^{(E)} + h_+^{(B)}$ with individual contributions from the fluid and the electric and magnetic fields:

{
\begin{subequations}
\label{eq:h+_fem}
\begin{align}
    h_+^{(matter)} &= \frac{2}{D} \int \diff^3\vec{r}\, (\varepsilon+P)\, \Gamma^2 \beta^2\, \mathcal{F}(\uv{\beta}),
    \label{eq:h+_fem:f}
    \\
    h_+^{(E)} &= - \frac{1}{2\pi D} \int \diff^3\vec{r}\, E^2\, \mathcal{F}(\uv{E}),
    \label{eq:h+_fem:e}
    \\
    h_+^{(B)} &= - \frac{1}{2\pi D} \int \diff^3\vec{r}\, B^2\, \mathcal{F}(\uv{B}),
    \label{eq:h+_fem:m}
\end{align}
\end{subequations}
where 
\begin{equation} 
\mathcal{F}(\uv{v}) \equiv \left( \uv{n}_1\cdot\uv{v} \right)^2 - \left( \uv{n}_2\cdot\uv{v} \right)^2
 \label{eq:F}
\end{equation} 
is the dimensionless polarization factor of a direction $\uv{v}$, and all the quantities $\varepsilon$, $P$, $\vec{\beta}$, $\vec{E}$, and $\vec{B}$ are functions of $\vec{r}$ and $t_r$.
}

Eqs.~\eqref{eq:h+_fem} illuminate two features of GW emission by a relativistic source. First, fluid and EM contributions have similar forms and can be transformed from one to another with a substitution 
\begin{equation}
    \label{eq:beta-E-B}
    \Gamma \vec{\beta} \sqrt{\varepsilon+P}
    \leftrightarrow 
    \frac{\uv{n}\times\vec{E}}{\sqrt{4\pi}}
    \leftrightarrow
    \frac{\uv{n}\times\vec{B}}{\sqrt{4\pi}}.
\end{equation}

The second feature comes from the quadrupole nature of GW emission. A relativistic source, typically a jet, has a mean direction of propagation. Then for each piece of matter in the jet, longitudinal and transverse components of its motion contribute to $h_+$ with different signs. In case of a periodic GW signal, this effect would be barely noticeable as a phase shift, but for aperiodic memory-type signals it is prominent. In fact, this means that for a jet with no EM component, a change of sign of its GW emission is evidence for a change of the ``GW-brightest'' component, from longitudinal to transverse or vice versa. The same is true for a magnetized jet, with an addition that the longitudinal components of $\vec{E}$ and $\vec{B}$ contribute to GW emission with the same sign as the transverse fluid motion, and, vice versa, transverse parts of $\vec{E}$ and $\vec{B}$ contribute with the same sign as the longitudinal fluid propagation.

\begin{figure}[t]
  \centering
  \includegraphics[width=0.5\textwidth]{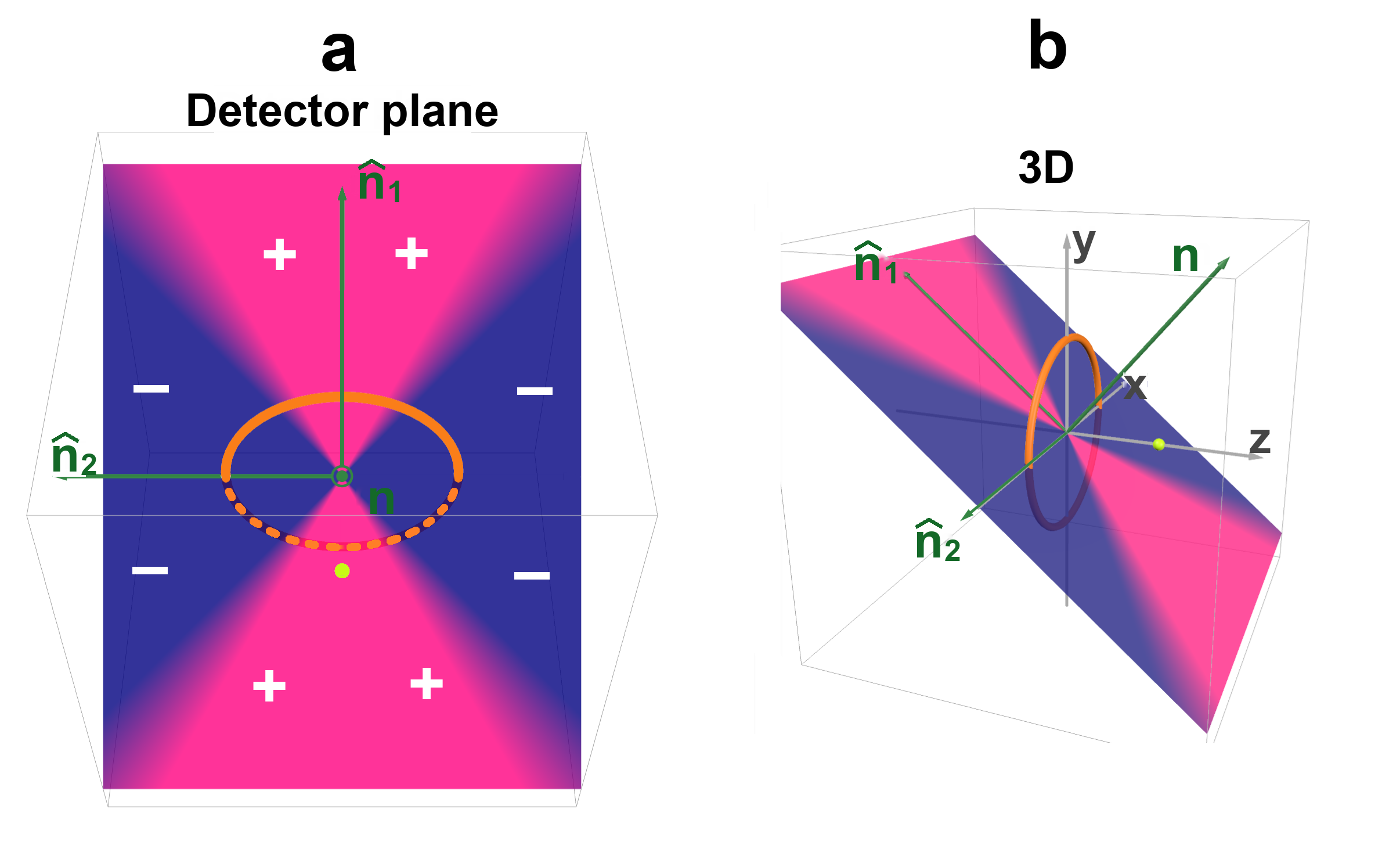}
  \caption{Contribution of each velocity direction to $h_+$.
  \textbf{(a)} The detector plane, viewed along the observer direction
$\uv{n}=\cos\theta_v\,\uv{z}+\sin\theta_v\,\uv{y}$,
  where $\uv{z}$ is the jet axis. The plane is spanned by
  $\uv{n}_1$ and $\uv{n}_2$.
  \textbf{(b)} The same plane shown in 3D.
  The colour shows the sign of ${\cal F}(\uv{\beta})$ (Eq.~\ref{eq:F}) for a
  velocity direction whose projection falls at that point: pink for positive and
  blue for negative.
  The orange ellipse is the set of projections of the radial and azimuthal unit
  vectors over all angles around the jet axis; it lies mostly in the negative region.
  The yellow point is the projection of the axial unit vector $\uv{z}$,
  which lies in the positive region.}
  \label{fig:5}
\end{figure}

{
This sign rule is illustrated in Fig.~\ref{fig:5}. A velocity direction
$\uv{\beta}$ contributes to $h_+$ with the factor $
{\cal F}(\uv{\beta})
=(\uv{n}_1 \cdot \uv{\beta})^2-(\uv{n}_2\cdot \uv{\beta})^2
=\sin^2\theta_\beta\,\cos 2\varphi_\beta $,
where $\theta_\beta$ is the angle between $\uv{\beta}$ and the
observer direction $\uv{n}$, and $\varphi_\beta$ is the azimuthal
angle of its projection onto the detector plane, measured from
$\uv{n}_1$ (see Fig.~\ref{fig:Vector_basis}). Hence ${\cal F}$ depends only on the projection of
$\uv{\beta}$ onto the detector plane: it is positive when the
projection lies closer to $\uv{n}_1$ and negative when it lies
closer to $\uv{n}_2$.

The axial direction $\uv{z}$ makes an angle $\theta_v$ with the line of sight, so its sky-plane projection is always aligned with $\uv{n}_1$ ($\varphi_\beta=0$) and falls in the positive region, with ${\cal F}(\uv{z})=\sin^2\theta_v>0$. Axial motion therefore makes a positive contribution to $h_+$.

For an axisymmetric source, the radial and azimuthal directions sweep through all angles
around the jet axis.  Projected onto the sky plane, they  form an ellipse with a semi-minor axis
$\cos\theta_v$ along $\uv{n}_1$ and a semi-major axis $1$ along
$\uv{n}_2$. 
Since the ellipse is squeezed along $\uv{n}_1$, most of it falls in the negative region, and averaging over it gives $\langle {\cal F}\rangle=\tfrac12(\cos^2\theta_v-1)=-\tfrac12\sin^2\theta_v<0$. Radial and azimuthal motions therefore make a negative contribution to $h_+$.
}

The same geometric argument applies to the electromagnetic contributions,
Eqs.~\eqref{eq:h+_fem:e} and ~\eqref{eq:h+_fem:m}, with $\uv{\beta}$ replaced by the direction of
$\vec{E}$ or $\vec{B}$. Because of the overall minus sign in these
equations, the signs are reversed: the axial components of $\vec{E}$ and
$\vec{B}$ contribute negatively to $h_+$, while their radial and azimuthal
components contribute positively.

It is important to note that these sign rules are purely geometric: they follow from the orientation of the jet axis relative to the detector plane and hold for every viewing angle. The viewing angle changes only the magnitude of each contribution, through the factors $\sin^2\theta_v$ and $-\tfrac12\sin^2\theta_v$, never its sign. As we show later, this allows us to infer the inner activity of the GRB engine, which cannot be probed otherwise.

\section{Methods}
\label{sec:methods}

Our goal is to compute, following a numerical simulation of a jet, the signal obtained from Eq.~\eqref{eq:h+_fem} for a given source that contains both matter and EM contributions. We consider a cylindrical geometry; however, the scheme and method developed here can be easily extended to three dimensions.
The integration that defines $h_+(t')$ is performed over a 3-dimensional surface in 4-dimensional spacetime.

Integrating at a fixed observer time $t'$, i.e., computing $h_+(t')$ over a surface of constant retarded time, can be challenging, as it requires holding {in memory} all possible time steps of the source configuration.
To address this issue, we perform the integration in reverse order. We divide the observer time into bins, each of which collects gravitational contributions. We then load the source configuration at a fixed time step and, for each occupied cell, add its gravitational contribution (a ``numerical graviton'') to the bin corresponding to the arrival time of that emission. Summing these contributions over all available laboratory times yields the resulting signal.
However, it is essential to determine which observer-time bins are fully populated. When parts of the jet move relativistically toward the observer, material at very late laboratory times (which may not be available in the data) can get so close to the observer that its emission falls into very early bins. This is one of the main challenges in numerically estimating the final signal, and it can render the signal incomplete at early observer times. In the results section, the curves for small viewing angles, where the jet material propagates toward the observer, are therefore truncated at early times.

In practice, defining the bins amounts to assigning a finite thickness $\Delta t'$ to the equal-arrival-time surfaces, which allows us to replace the exact integral with a summation over a 4-dimensional lattice. The lattice is the Cartesian product of the 3-dimensional spatial grid, with cells of size $\Delta\phi\,\Delta R\,\Delta z$, and the laboratory time grid, whose points are integer multiples of the laboratory time step $\Delta t$.

Therefore, the quantity  computed 
is the time-averaged signal over an observer-time window of width $\Delta t'$:
\begin{equation}
\begin{aligned}
    \bar{h}_{m'} := \bar{h}(t') 
    &= \frac{1}{\Delta t'} 
    \int_{t' - {\Delta t'}/{2}}^{t' + {\Delta t'}/{2}} 
    h(s') \, \diff s' 
    \\
    &= \sum_{(i,j,k,m)\in I_{m'}} \delta h_{ijkm}
\end{aligned}
\label{eq:expression_for_h_m'}
\end{equation}
where $t' := \Delta t' m'$, and $m' \in \mathbb{Z}$ is the observer-time index corresponding to $t'$ .
We denote by $I_{m'}$ the set of all spacetime-cell indices $(i,j,k,m)$  whose centers lie within this ``equal-observer-time volume'' $V_{m'}$.
\begin{equation}
\begin{aligned}
I_{m'} 
= \Big\{ (i,j,k,m) \;\Big|\; 
&\Delta t \!\left(m-\tfrac{1}{2}\right) 
\le 
\Delta t' m' +\vec{r}_{ijk} \cdot \uv{n} 
\\
&\le 
\Delta t \!\left(m+\tfrac{1}{2}\right)
\Big\},
\end{aligned}
\label{eq:observer_time_4d_volume}
\end{equation}
and 
$\Delta t' m' + \vec{r}_{ijk} \cdot \uv{n}$ corresponds to the retarded time. 
The quantity $\delta h_{ijkm}$
denotes the contribution from a single spacetime lattice cell: 
\begin{equation}
    \begin{aligned}
    \delta h_{ijkm} 
    &= \frac{2 \Delta t \, \Delta V_{ijk}}{D}
    \Bigg[
        \beta_{ijkm}^2 \Gamma_{ijkm}^2 
        \\
         \times &(\varepsilon_{ijkm}+P_{ijkm}) \,
        {\cal F}(\uv{\beta}_{ijkm})
        \\
       -& \frac{1}{4\pi}
        \left(
            E_{ijkm}^2 \, {\cal F}(\uv{E}_{ijkm})
            + B_{ijkm}^2 \, {\cal F}(\uv{B}_{ijkm})
        \right)
    \Bigg],
    \end{aligned}
    \end{equation}

\begin{figure}
  \centering
  \includegraphics[width=\columnwidth]{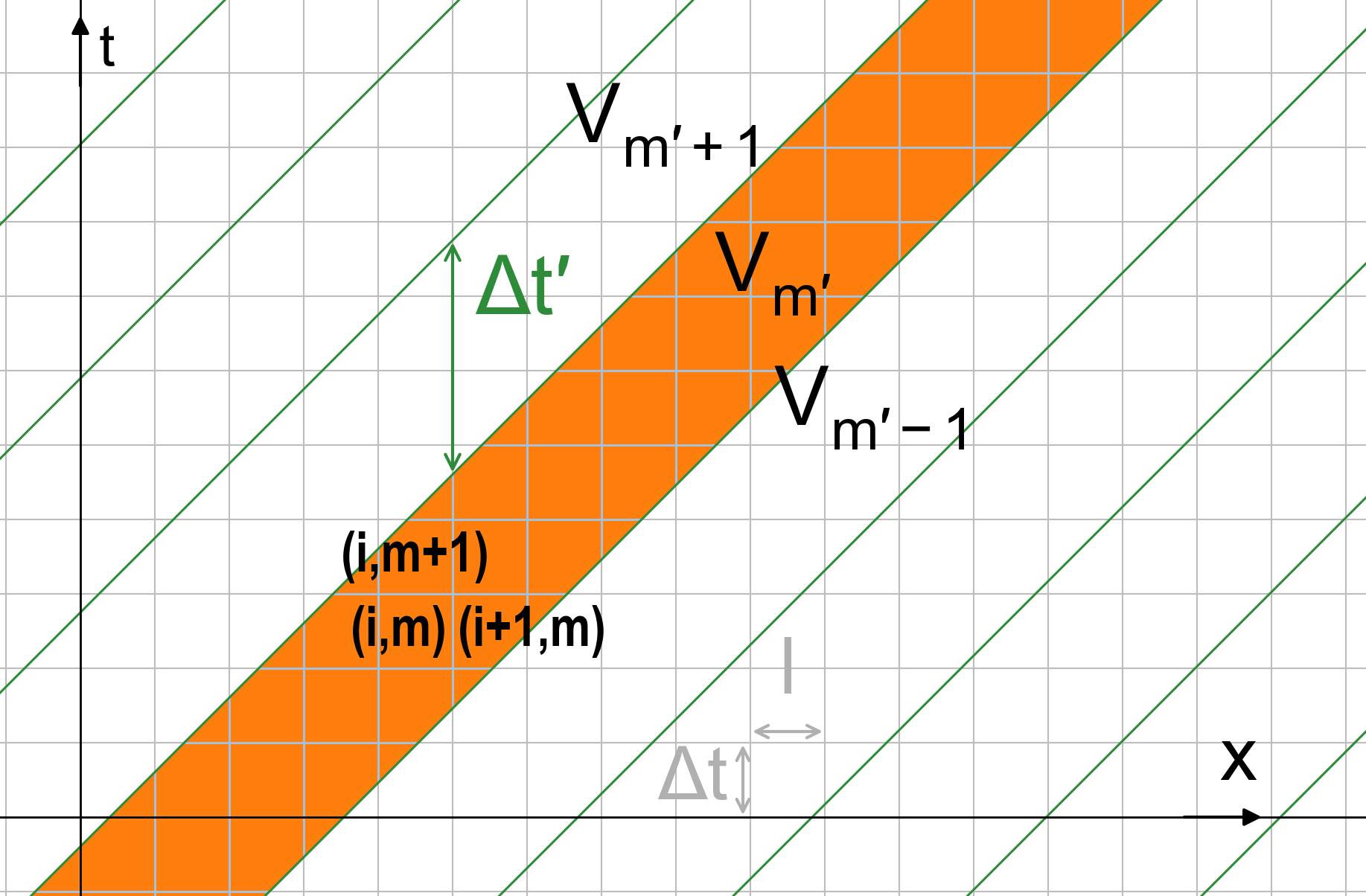}
  \caption{Two-dimensional spacetime discretized into a lattice. The horizontal spacing corresponds to the spatial grid size $l$, and the vertical spacing corresponds to the laboratory timestep $\Delta t$. The orange strip represents the spacetime region $V_{m'}$ contributing to $\bar h_{m'}$. The separation between the green lines (surfaces) is the observer-time window $\Delta t'$.}
  \label{fig:lattice_contribution_errors}
\end{figure}

The discrete version of the integration in Eq.~\eqref{eq:expression_for_h_m'} is illustrated in Fig.~\ref{fig:lattice_contribution_errors}. Each spacetime lattice cell $(i,m)$ in the figure represents a spacetime element that can emit gravitational radiation. To compute $\bar h_{m'}$, we sum the contributions from all spacetime lattice cells whose centers fall inside the ``{equal-}observer-time volume'' $V_{m'}$, shown as the orange strip in the figure. In our simulation, each spacetime cell is defined as the Cartesian product of a 3D cylindrical spatial cell and a time interval 
{$[(m-\frac{1}{2})\Delta t, \, (m+\frac{1}{2})\Delta t]$}.

{
The retardation effectively splits what would otherwise be a single, sequential three-dimensional summation into many smaller two-dimensional summations, each contributing to a different observer-time bin. This allows us to divide the computation into independent processes that can run in parallel.
Furthermore, the computation of the integrand itself in Eq.~\eqref{eq:h+_fem},
involves many vector operations, which can be executed in parallel and are naturally optimized for fast execution on GPU cores. These advantages make the use of GPU kernels beneficial. 
Additionally, it is always possible to decompose a chronological summation into hierarchical iterations of smaller independent summations, reducing synchronization overhead and further boosting performance.
}

{
Our code uses a uniform cylindrical grid with fixed $\Delta z$, $\Delta r$, and $\Delta\phi$, whereas the MHD simulation data are defined on a non-uniform spherical grid that spans four orders of magnitude and captures both the jet and the counter-jet over the available time. The cell size ranges from $\sim0.014\,R_{\rm in}\approx33$\,km in the inner region to $\sim400\,R_{\rm in}\approx10^{11}$\,cm in the outer region. To handle this, we run the computation in three iterations, on cylindrical domains of size $48\times24$, $768\times384$, and $12288\times3072$ (length $\times$ radius, in units of $R_{\rm in}$). In each run, we exclude the region already covered by the previous, smaller domain to avoid double counting. Each run uses $2048\times1024\times384$ axial, radial, and angular divisions (about one billion cells), so that our cells are no larger than the overlapping MHD cells in every regime. We process all $10^4$ time steps available in the simulation data ($\Delta t = 4$\,ms) and set the observer-time bin width similarly.
}

\section{Code tests and generic features of GW from jets}
\label{sec:test}

We apply the numerical scheme described above to a sequence
of jet models of increasing physical complexity. We begin with analytically
controlled spherical-cap models, which serve as tests of the code and illustrate
the basic temporal and angular dependence of jet-GW signals.

\begin{figure}
  \centering
  \includegraphics[width=0.85\columnwidth]{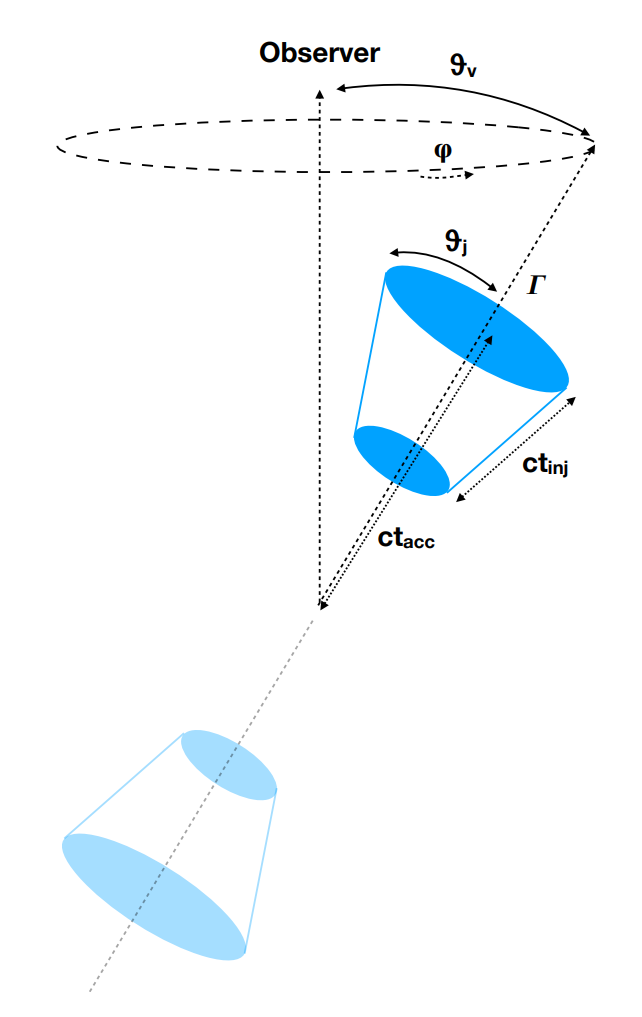}
  \caption{Schematic description of the cap model with a continuous injection. The leading shell
  reaches its final Lorentz factor at a distance $ct_{\rm acc}$ from the origin.
  The duration of mass injection is $t_{\rm inj}$. From \citet{leiderschneider2021}.
  Also shown are the viewing angle $\theta_v$ and the jet angle $\theta_j$.
  A counter-jet is shown in
  light colors. 
  }
  \label{fig:cim_jet_model}
\end{figure}

We test the code using a thin spherical cap with an opening angle
$\theta_j$. The cap consists of particles moving radially outward, and the
observer is located at an angle $\theta_v$ relative to the cap axis (see Fig.~\ref{fig:cim_jet_model}). The
retarded time $t_r$ satisfies $t_r={t'} +\cos\theta\,R(t_r) $,
where $\theta$ is measured relative to the observer direction. Using the
point-particle expression, Eq.~\eqref{eq:h+_fem:f}, the
total plus-polarized signal from the cap is
\begin{align}
h_+({t'})
&=
\frac{2 M_{\Omega}}{D}
\int_{\max(0,\theta_v-\theta_j)}^{\min(\pi,\theta_v+\theta_j)}
\diff\theta
\int_{-\Delta_\phi(\theta)}^{\Delta_\phi(\theta)}
\sin\theta\,\diff\phi
\nonumber\\
&\quad\times
\left[
\frac{\sin^2\theta}{1-\beta(t_r)\cos\theta}
\Gamma(t_r)\beta^2(t_r)
\cos(2\phi)
\right]
\nonumber\\
&=
\frac{2M_{\Omega}}{D}
\int_{\max(0,\theta_v-\theta_j)}^{\min(\pi,\theta_v+\theta_j)}
\diff\theta\,
\frac{\sin^3\theta}{1-\beta(t_r)\cos\theta}
\nonumber\\
&\quad\times
\Gamma(t_r)\beta^2(t_r)
\sin\!\left[2\Delta_\phi(\theta)\right].
\label{eq:Spherical Integration Of PP GW}
\end{align}
Here $    \Delta_\phi(\theta)=
    \cos^{-1}[ (\cos\theta_j-\cos\theta_v\cos\theta)/
    {\sin\theta_v\sin\theta}]$ and
$    M_\Omega ={M_{\rm cap}}/{2\pi(1-\cos\theta_j)} $
is the rest mass per unit solid angle. One can notice here a prominent feature. The signal is \emph{anti-beamed}:
The amplitude vanishes along the direction of motion and is suppressed within a
cone of opening angle $\sim\Gamma^{-1}$ around it, so that in a realistic GRB jet an observer who is exactly along the axis and observes the maximal 
$\gamma$-rays signal is poorly placed to see the gravitational waves. The radiated
GW energy is beamed in the forward direction, but just outside the jet's
cone.

\begin{figure}
  \centering
  \includegraphics[width=\columnwidth]{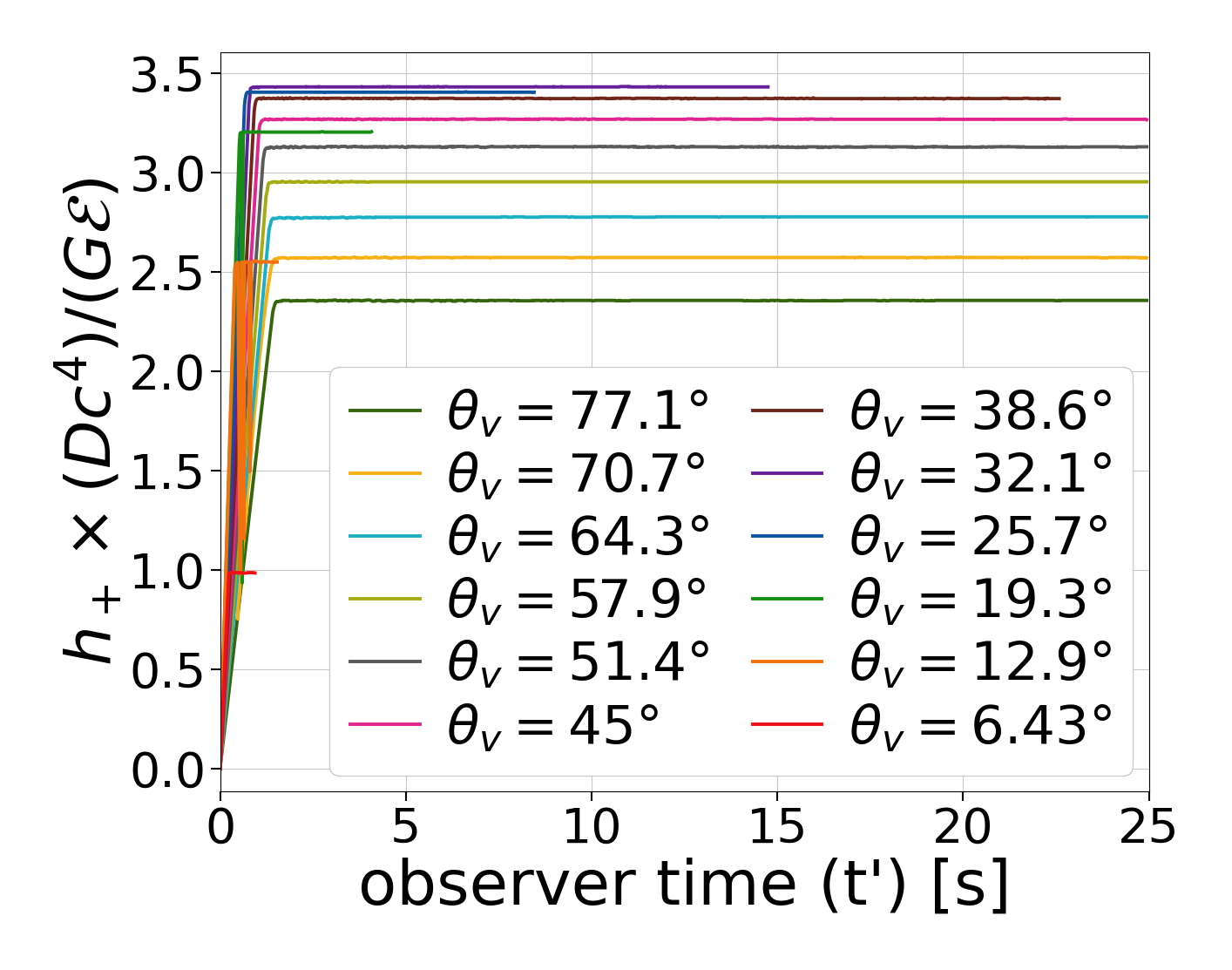}
  \includegraphics[width=\columnwidth]{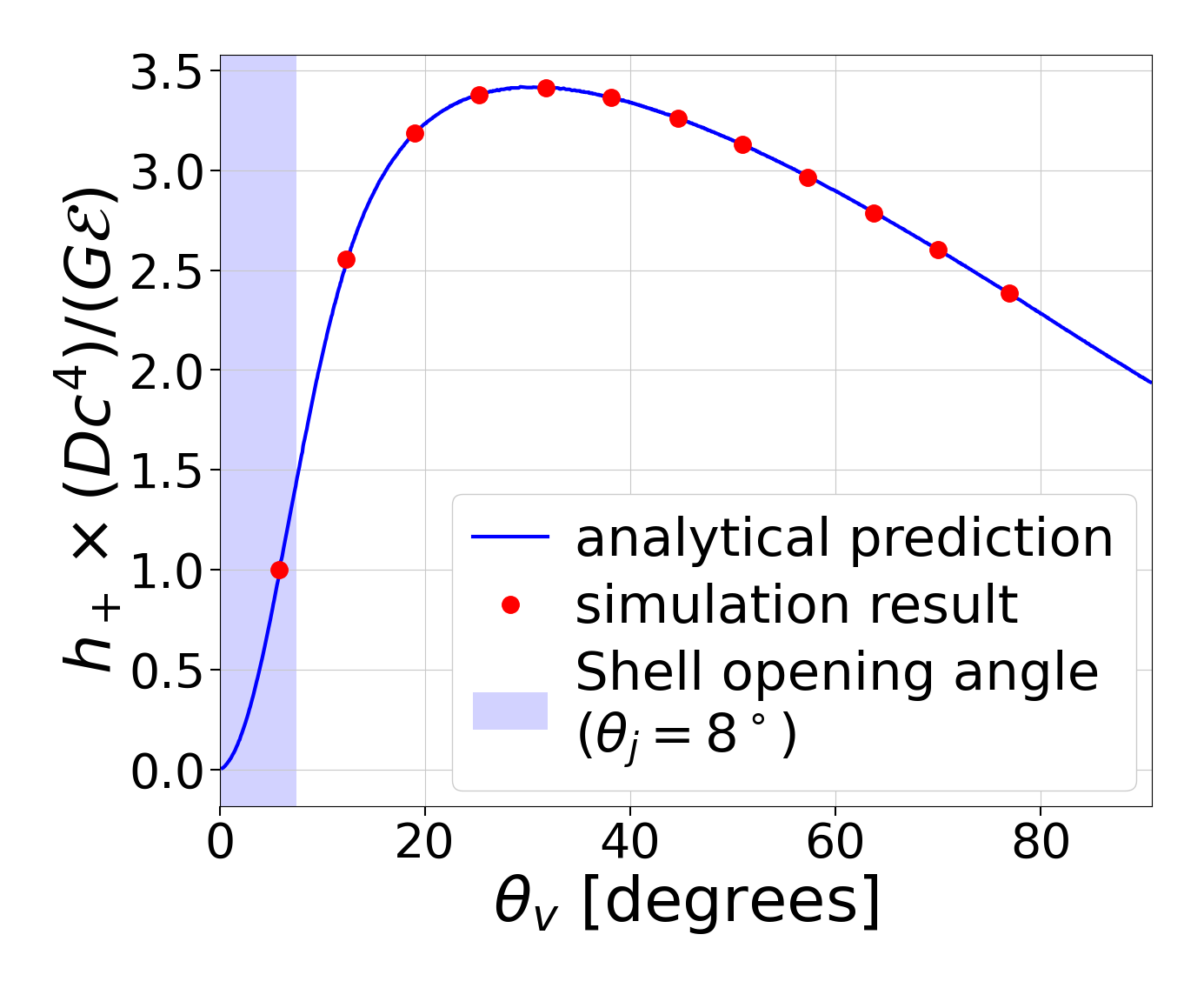}
  \caption{{\it Top:} The plus mode of the gravitational-wave signal from an
  instantaneously accelerated spherical shell for different observer angles.
  The shell has opening angle $\theta_j=8^\circ$, Lorentz factor $\Gamma=10$,
  total energy {${\cal E}$}, and distance $D$. Each signal is truncated at
  the first observer time affected by the shell's exit from the computational
  domain. {\it Bottom:} Comparison between the plateau values obtained
  numerically and Eq.~\eqref{eq:Theta Integral For h_+ For Instantaneous Acceleration}.
  The red points are numerical results, while the blue curve is the analytic
  prediction.}
  \label{fig:signal_from_instanteniuos_accelerated_shell}
\end{figure}
For instantaneous acceleration to a velocity $\beta_0$,
$    t_r({t'},\theta)={{t'}}/({1-\beta_0\cos\theta})$,
and Eq.~\eqref{eq:Spherical Integration Of PP GW} reduces to \citep{leiderschneider2021}
\begin{align}
h_+({t'})
&=
\frac{2M_\Omega\Gamma_0\beta_0^2}{D}\,
\Theta({t'})
\int_{\max(0,\theta_v-\theta_j)}^{\min(\pi,\theta_v+\theta_j)}
\diff\theta
\nonumber\\
&\quad\times
\frac{\sin^3\theta}{1-\beta_0\cos\theta}
\sin\!\left[2\Delta_\phi(\theta,\theta_v)\right].
\label{eq:Theta Integral For h_+ For Instantaneous Acceleration}
\end{align}
The observed signal is a memory-type step function. In the
ultrarelativistic limit, its amplitude is maximal near
\begin{equation}
    \theta_{\rm max}\simeq \sqrt{\frac{2}{\Gamma}} .
    \label{eq:expression_for_theta_max}
\end{equation}
In the numerical calculation we initialize the shell at a finite radius $r_{\rm inj}$, large enough to occupy a significant number of cells. This introduces an artificial rise time, since the observer sees the nearest part of the shell before the farthest part, and it also introduces a late-time cutoff when the shell leaves the computational domain.
Figure~\ref{fig:signal_from_instanteniuos_accelerated_shell} compares the numerical result with the analytical prediction. The agreement is excellent. The maximum signal occurs at $\theta_v\simeq 30^\circ$, close to the ultrarelativistic prediction $\theta_{\rm max}\simeq25.6^\circ$ for $\Gamma=10$.
One can note in this figure the time of flight effect. For smaller angles, the calculated signal extends to much shorter observer times.

\begin{figure}
  \centering
  \includegraphics[width=\columnwidth]{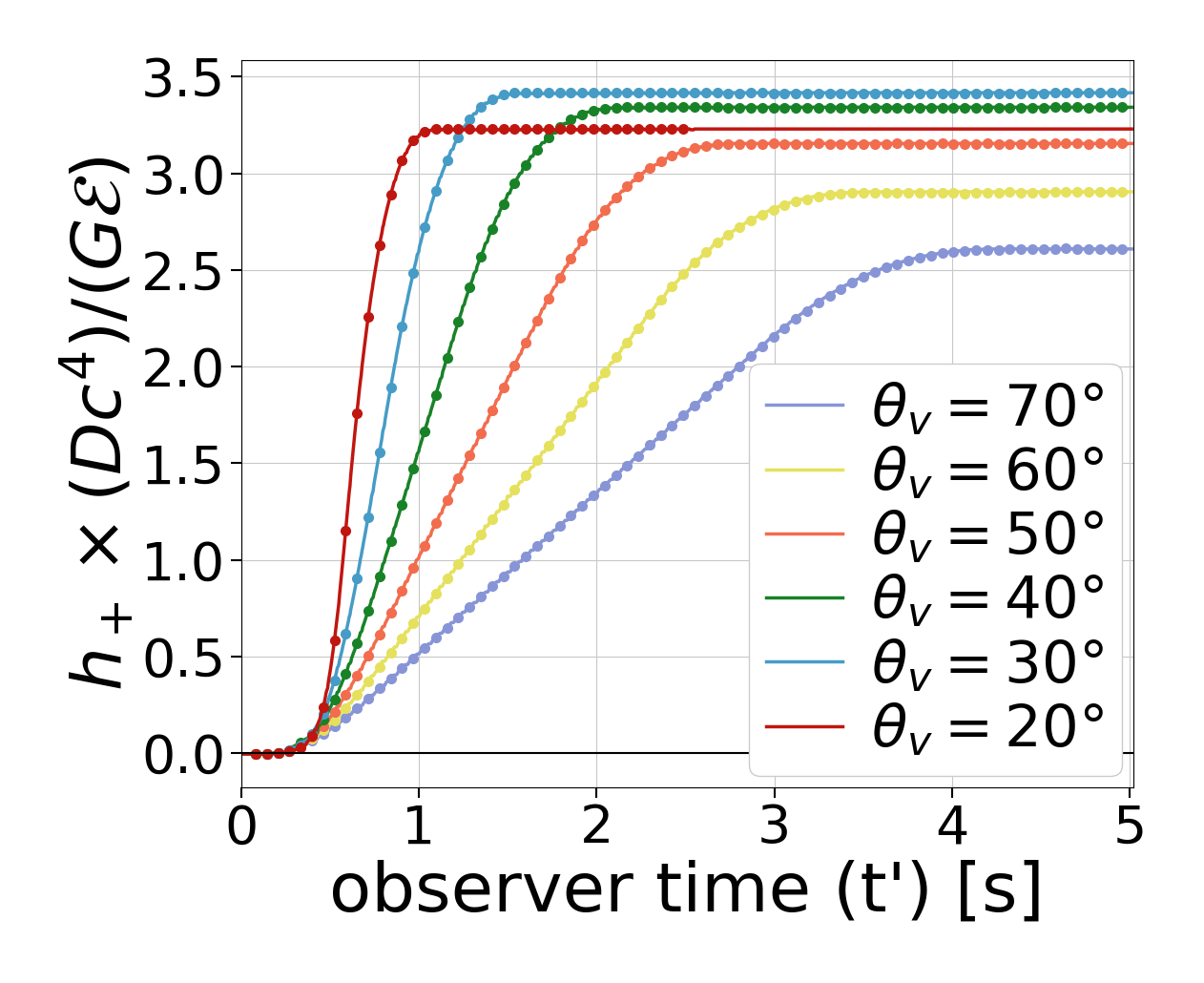}
  \caption{A comparison between the code results for a constantly accelerated
  shell and the {analytic} prediction obtained from
  {Eqs.~\eqref{eq:Spherical Integration Of PP GW} and~\eqref{eq:Expression for the constant acceleration retarded time}.} The
  parameters are $\theta_j=8^\circ$, $t_{\rm acc}=5\,{\rm s}$, and
  $\Gamma=10$.}
  \label{fig:constant_force_model_comparison}
\end{figure}

Figure~\ref{fig:constant_force_model_comparison} depicts a comparison between the
numerical calculation and a one dimensional integration of Eq.~\eqref{eq:Spherical Integration Of PP GW} that is possible after obtaining {an} analytic solution of the retarded time for this case.
Here, we considered a cap undergoing a constant (in the lab frame) acceleration,
$    {d}(\Gamma\beta)/{dt}=C $. In this case $R(t)=    (\sqrt{C^2t^2+1}-1)/{C}$, until it reaches $\Gamma_{\rm max}$.
    In this case, the retarded time is given by
    \begin{equation}
    \label{eq:Expression for the constant acceleration retarded time}
        t_r(t',\theta) = \frac{C t' + \cos\theta \left( \sqrt{C^2 t'^2 - 2C t' \cos\theta + 1} - 1 \right)}{C \sin^2\theta},
    \end{equation}
    which is substituted into Eq.~(10) to obtain the analytic result shown for comparison in Figure~5.
\begin{figure*}
  \centering
  \includegraphics[width=\textwidth]{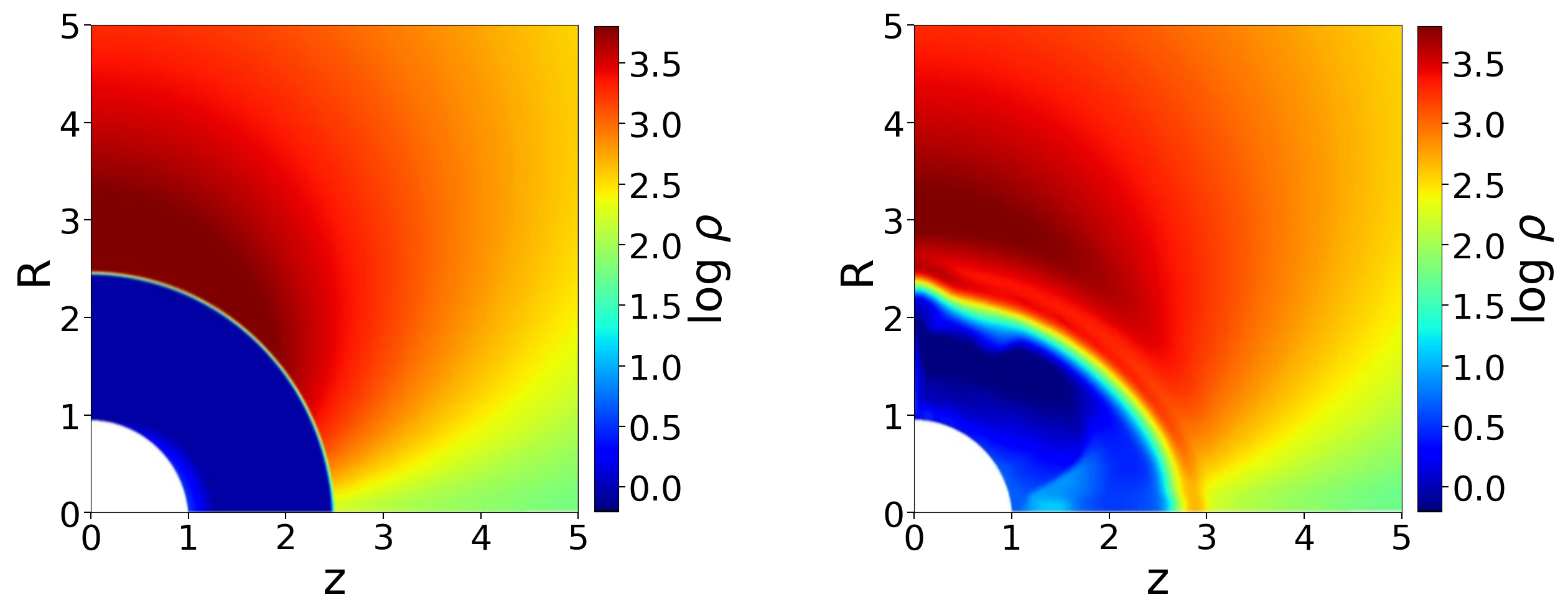}
  \caption{The baryonic matter density  at the inner part of the
  simulation region at the initial times, $t=0$ s
  {(left)}
  and shortly after the onset of energy injection $t=0.048$ s
  {(right)}.
  The jet axis $z$ is horizontal and the cylindrical radius $R$ is vertical. The white region corresponds to the conducting magnetized star.   The blue region (in the left figure) is the central cavity. One can notice that the cavity begins to expand in the z direction - the expected direction of the jet, but also a kick given early on to the matter along the equator. The corresponding radial matter motion is responsible for the initial negative GW signal. }
  \label{fig:density}
\end{figure*}
The shell has the same angular width as before, $\theta_j = 8^\circ$. 
The agreement between the analytic and the numerical result 
demonstrates that the code correctly captures the angular dependence,
retardation, and finite acceleration history of the source.  The results demonstrate again the expected angular dependence that is minimal along the shell's axis and increases up to a maximum that is slightly wider than $\theta_j$. Time-of-flight effects are observed here. At small angles, the observer sees a signal that rises much faster than the actual acceleration time. 

Similar satisfactory numerical tests have been obtained for a jet (rather than a thin cap) in which matter is continuously accelerated for a duration $t_{\rm inj}$. However, there is a significant difference here. While we see time-of-flight effects influence the relation between the acceleration time and the observed time they don't affect the injection time. An injection time of 
$ t_{\rm inj}$ is translated to a similar observer time.

\section{Jet Simulations}
\label{sec:jet}

To explore a more realistic configuration we used the 2D axisymmetric relativistic MHD simulation  of \cite{bromberg2018} of a jet propagating in a homologically expanding ejecta remnant of binary neutron star merger. The simulation begins $0.5$ s after the merger when the ejecta has already developed and the magnetic field of
the proto-neutron star has grown enough to launch a Poynting-dominated jet.

\begin{figure*}
    \centering
    \includegraphics[width=0.87\textwidth]{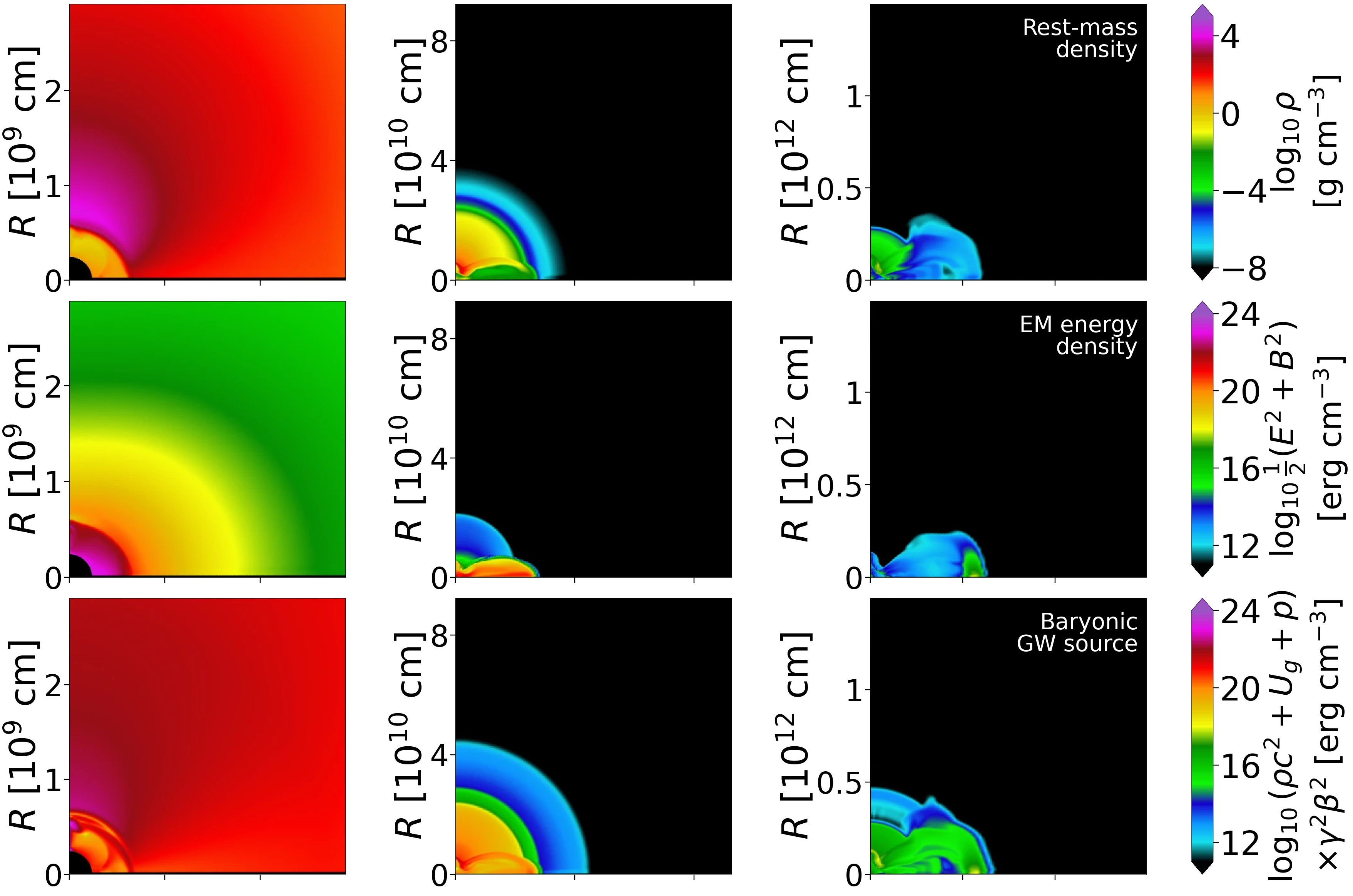}\\[2mm]
    \includegraphics[width=0.863\textwidth]{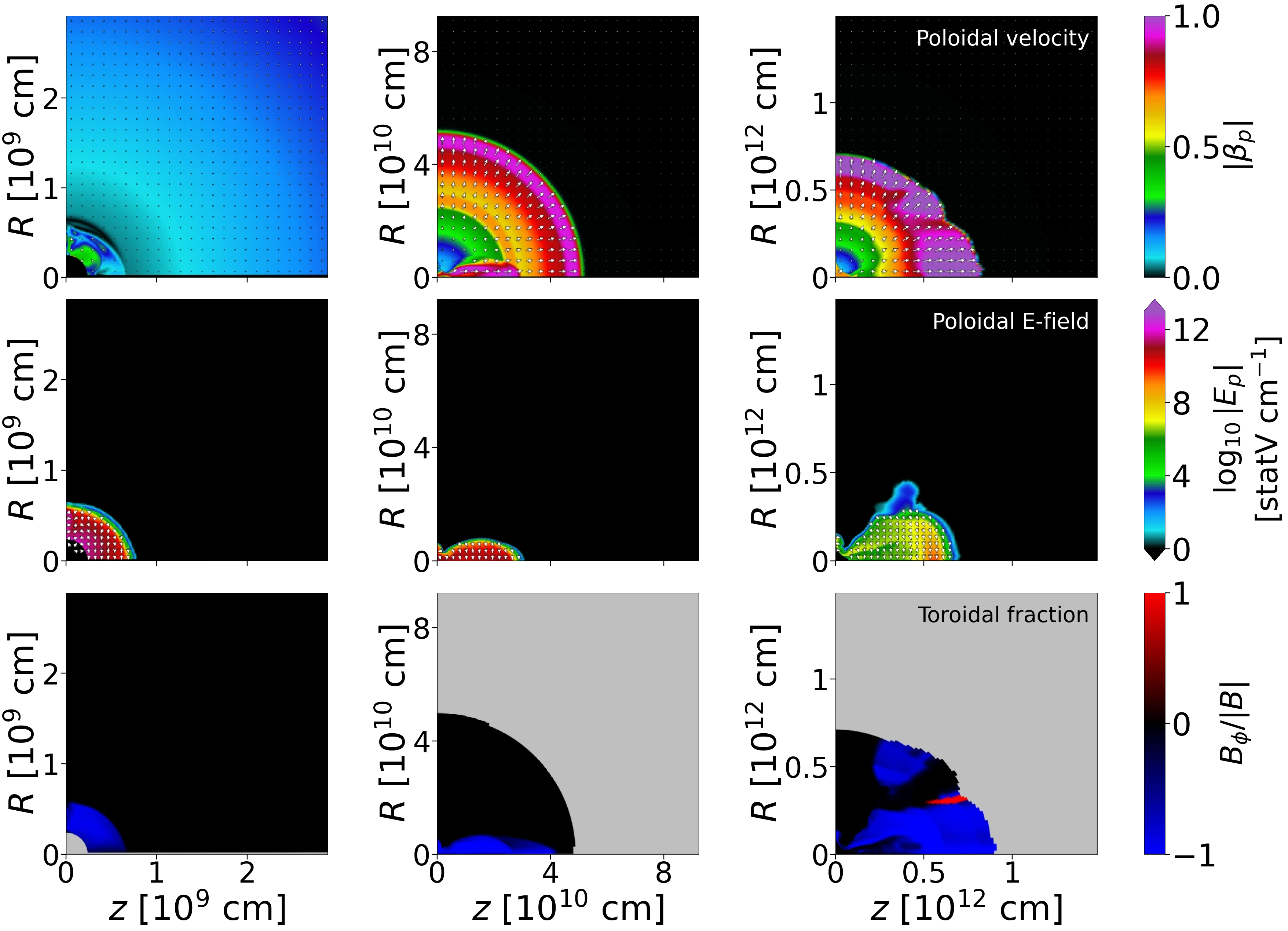}\hspace{1.2mm}
    \caption{Overview of the MHD jet evolution at,
  $t = 0.04$, $1.2$ and $20$ s (left, center and right).
 From top to bottom: the baryonic rest-mass density $\rho$;
  the electromagnetic energy density $(E^2+B^2)/8\pi$;
  the baryonic GW source term, i.e.\ the integrand of Eq.~\eqref{eq:h+_fem:f};
  the velocity $\beta$; the electric field $E$; and the toroidal magnetic
  field $B_\phi$. The left column shows the initial interaction of the
  Poynting flux with the inner cavity and the surrounding ejecta, the center
  column the Poynting-flux-dominated jet during the injection phase, and the
  right column the mostly baryonic jet after it has broken out of most of
  the ejecta. }
    \label{fig:overview1}
\end{figure*}
{The ejecta follows the density and velocity profiles of \cite{Hotokezaka2013},
made spherically symmetric: total mass about $0.06 m_\odot$, extending from
$7.5 \times 10^6$ cm to $6 \times 10^9$ cm,
with a homologous 4-velocity profile reaching $0.4$ c at the outer edge.
The ejecta is surrounded by a light, fast tail of $10^{-7} m_\odot$ with a
density falling as $r^{-12}$, and a homologous 4-velocity profile peaking at
$0.85$ c at $1.2 \times 10^{10}$ cm. At the center of the grid we place a
perfectly conducting sphere with a radius $R_s = 3\times10^6$ cm threaded by a
dipole field, representing the proto-neutron star. At $t=0$ the sphere is
rotated at an angular velocity $\Omega=0.8\,c/R_s$, which sets the light
cylinder radius at $R_L = 1.25 R_s$, between the surface of the sphere and the
inner ejecta edge at $2.5 R_s$. The space between the proto-neutron star
surface and the inner ejecta edge is filled with floor density with a
negligible amount of mass.}

The rotation winds the dipole field creating a toroidal magnetic component and an associated meridional electric field, the cross product of which generates Poynting flux in the radial direction. Field lines crossing the light cylinder open up and carry that flux into the jet, with the open-field polar angle scaling as $\sqrt{R_s/R_L}$, so the power follows the standard dipole-loss formula. 
{The process of field lines opening up generates an energy pulse in a direction perpendicular to the jet axis, which compresses material producing a density wave in the same direction (see Fig. \ref{fig:density}}).
The field strength at the inner boundary is chosen so that the electromagnetic luminosity is $4 \times 10^{50}$ erg/s per jet, equivalent to a millisecond proto-neutron star of 12 km radius with a surface dipole field of about $4 \times 10^{15}$ G. 
{At $t_{\rm inj} = {2\,{\rm s}}$ we stop the rotation and {cut off} the energy injection into the jet. As a result the jet is disconnected from the launching point and {continues} to propagate under its own inertia.}

\begin{figure}
    \centering
    \includegraphics[width=1.0\linewidth]{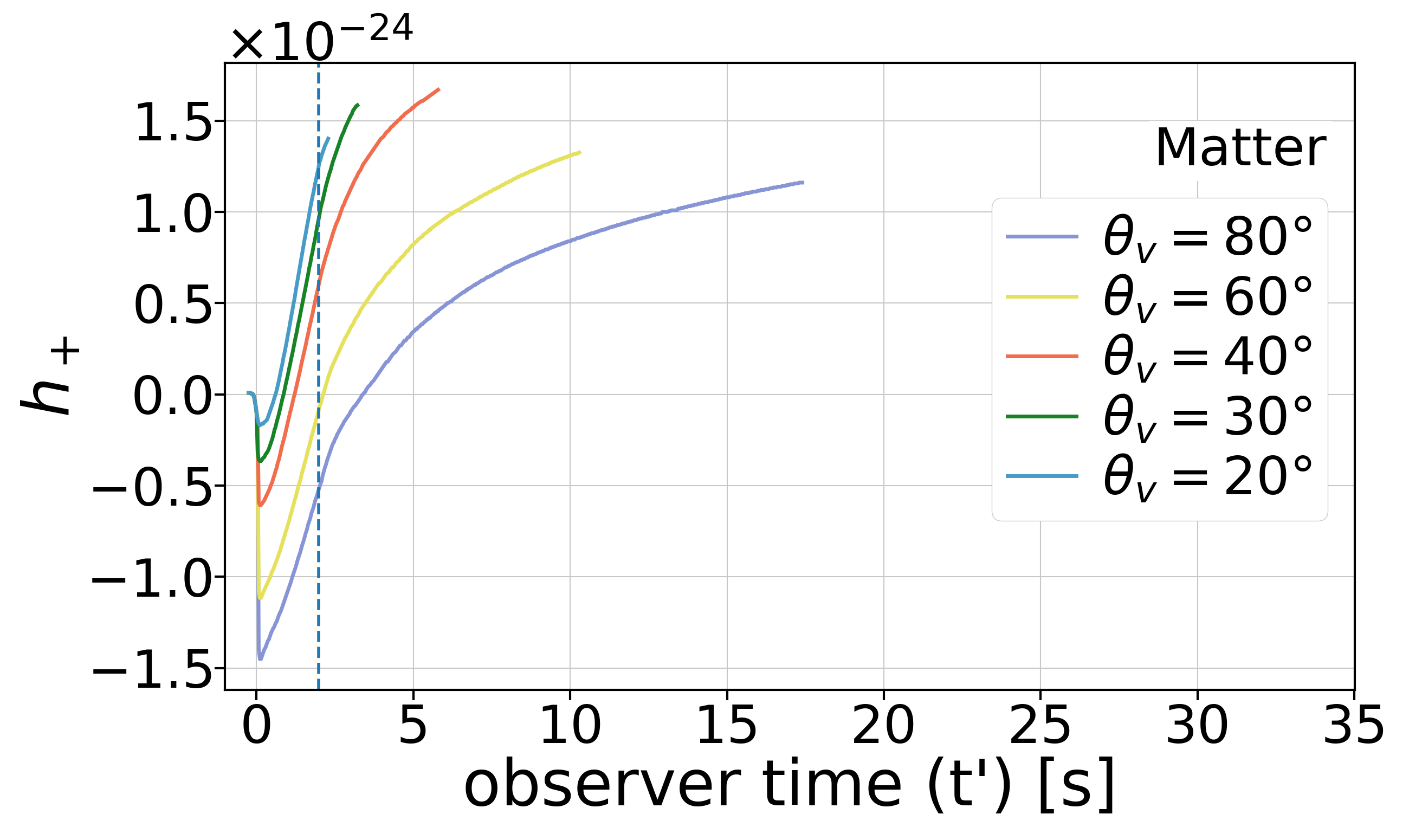}
    \includegraphics[width=1.0\linewidth]{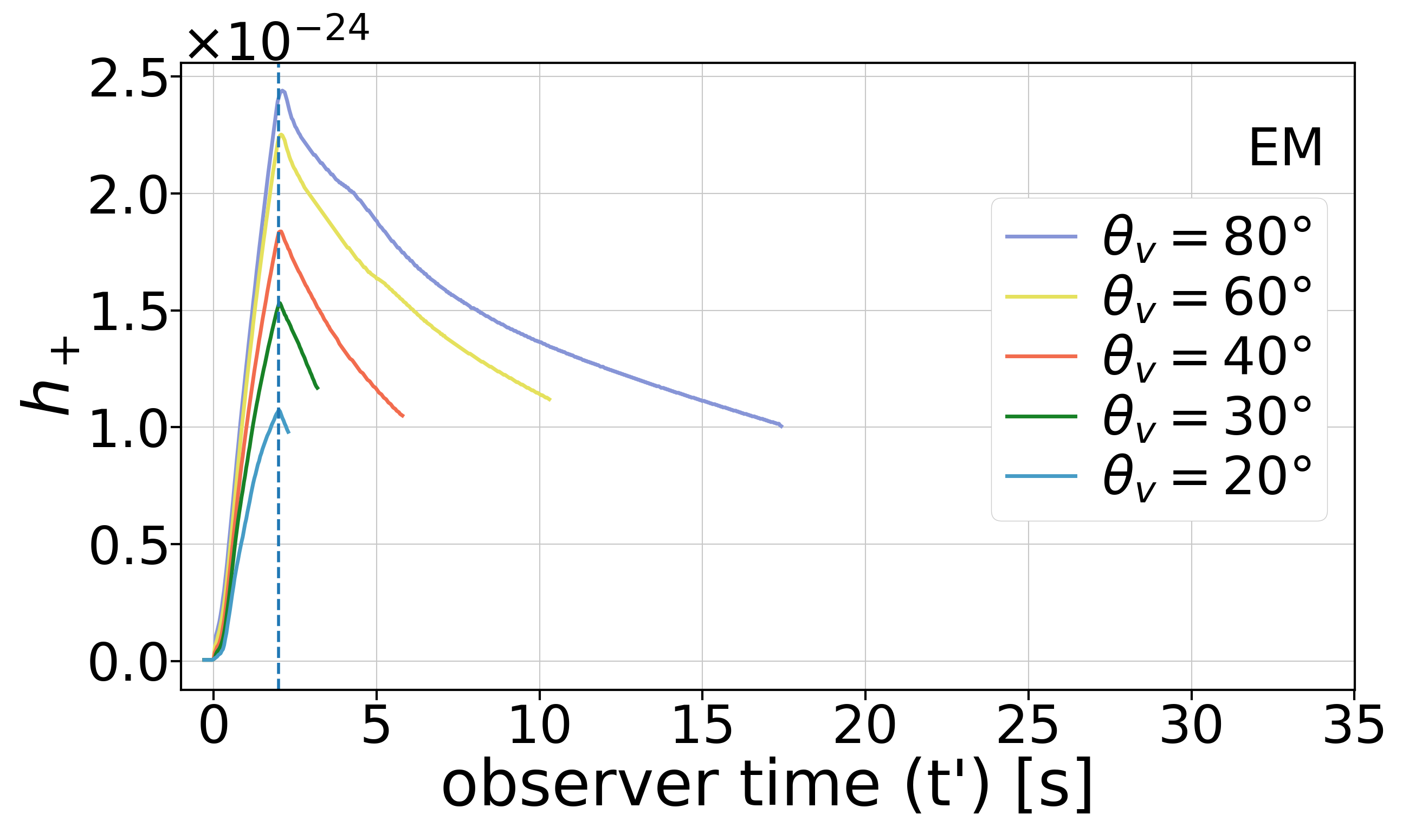}
    \includegraphics[width=1.0\linewidth]{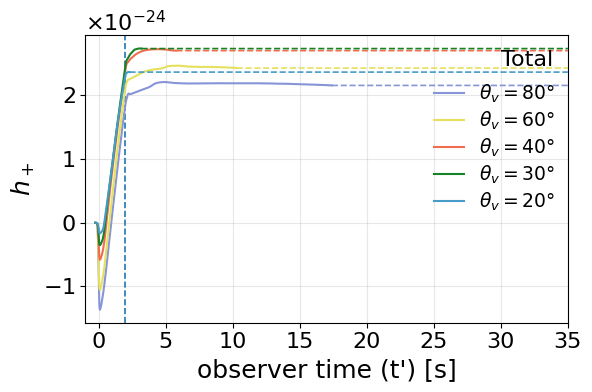}
    \caption{The GW waveforms from a {$4 \times 10^{50}$ erg/s} jet with a duration of {$t_{\rm inj} = 2$ s}, and the source distance is
   $D = 10\ \mathrm{Mpc}$. The {dashed line marks $t' = t_{\rm inj}$, the observer time at which the end of the energy injection is seen.}
    {\it top:} the matter contribution, {\it middle:} the EM contribution, {\it bottom:} the total signal. The dashed horizontal lines at the bottom panel, reflect the extrapolated constant signal after the truncation time. An interesting feature is the very early negative matter signal. This arises at the moment that the Poynting flux hits the inner boundary of the central cavity (see Fig.~\ref{fig:density}). At this moment the dominant matter motion is perpendicular to the final jet direction, leading to a negative contribution. This is an indication of the ability of the GW signal to explore details of the inner motion of the GRB engine. At small angles the signals are truncated because of the retardation effect.   }
    \label{fig:waveform}
\end{figure}
Fig. \ref{fig:overview1} {depicts} the overall behavior of the matter and EM at three time steps. The first is just when Poynting flux begins to be injected into the system, when one can clearly see the inner cavity and the initial interaction of the magnetic flux and the surrounding matter. At the second column one can see clearly the Poynting flux jet. The third column depicts a later time scale and a much larger scale. At this stage the jet is mostly baryonic and it has broken out from most of the ejecta.

\section{The Gravitational Waves }
\label{sec:results}

Using the GW code discussed above we now postprocess the output of the MHD jet simulations. 
Fig. \ref{fig:waveform} depicts our main results. Shown are the GW waveforms that arise from the matter contribution, from the EM contribution and the total signal. The jet injects electromagnetic energy to the system, as clearly seen in
Fig.~\ref{fig:overview1}. The electromagnetic contribution begins at the onset
of the jet activity and increases during the injection phase. Eventually, after 2 s no
additional electromagnetic energy is injected into the system. The existing EM energy is partially converted into baryonic
(kinetic energy). Thus the eventual drop of the EM contribution to the GW signal. This is compensated by the increasing contribution of the matter. Throughout the event,
the signal of the electromagnetic channel is positive, corresponding to mostly
radial and azimuthal fields. 

In the calculations we include the contribution of two jets. The one that is going towards the observer and the counter jet that is pointing away. For small angles the contribution of the counter jet is negligible. However, as expected \citep{leiderschneider2021}, for large angle the signal is almost doubled.

{As a sanity check, the plateau of the total signal, $h_+ \simeq 2.5 \times 10^{-24}$ at $\theta_v = 30^\circ$--$40^\circ$, can be compared with the point-source memory amplitude quoted in \S\ref{sec:intro}. For the energy injected into one jet, ${\cal E} = L\, t_{\rm inj} = 8 \times 10^{50}$ erg, and $D = 10$ Mpc, the prefactor $2G{\cal E}/(c^4 D)$ is $4.3 \times 10^{-24}$, and the angular factor $\beta^2 \sin^2\theta_v/(1-\beta\cos\theta_v)$ is at most $\simeq 1.8$ at these angles. The simulated plateau is thus about a third of the ultrarelativistic point-source limit, as expected since only part of the injected energy ends up in relativistic motion along the jet axis, the rest being deposited in the mildly relativistic cocoon.}

\begin{figure}
    \centering
  \includegraphics[width=1.0\linewidth]{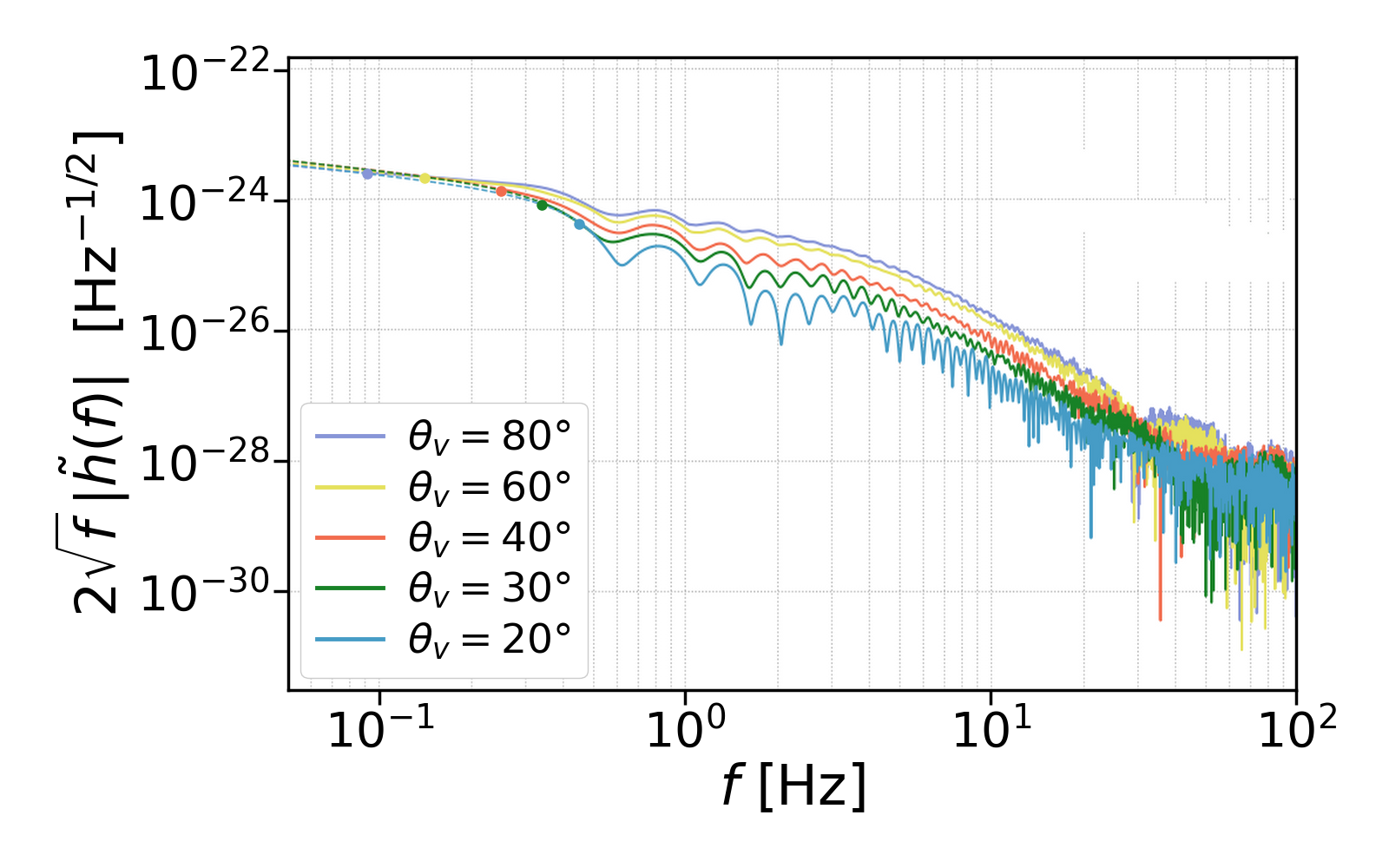}  
    \caption{Amplitude spectral density (ASD) of the  gravitational-wave
  signal. The dots mark the frequencies at which the calculation is truncated  and the signal is extrapolated as a constant (see Fig.~\ref{fig:waveform}). At large angles the expected $f^{-1/2}$ behaviour is observed even above these frequencies. Because of the temporal resolution there is an upper limit  at the Nyquist frequency $f=125$ Hz. This is slightly above the upper frequency range shown. }
    \label{fig:asd}
\end{figure}

Several clear features emerge. First we see the clear rise of the signal over 2 s. This is $t_{\rm inj}$, the duration of energy injection,  discussed earlier. Retardation effects don't influence this time scale that consequently doesn't depend on the viewing angle. This will be also reflected in the spectral density by a  corresponding {knee at $\sim 0.3$--$0.5$ Hz, of order $1/t_{\rm inj}$} (see Fig.~\ref{fig:asd}). 

Second, we still see the retardation (or time-of-flight) effect as for smaller angles the signal is truncated at earlier times. For these  observer times some of the source contribution material reaches the edge of the grid and its further contribution is not calculated. At the same time, because of this retardation effect  for the highest frequencies the strongest signals are obtained for the smaller viewing angles. In Fig~\ref{fig:waveform} we extrapolate the signal beyond this observer time as a constant value. This extrapolation is marked by horizontal dashed lines. It is expected to hold until the jet is slowed down by interaction with external matter, the interaction that produces the afterglow. We expect $h_+$ to decrease and gradually approaches zero. The corresponding observer time will depends very strongly on the viewing angle but even for small observer angles it is expected to be more than a dozen seconds.

Third, probably the most interesting and unexpected result is the negative peak observed at early times. The global minimum is reached at $t' \approx 0.044\ {\mathrm{s}}$. This peak appears when the Poynting flux emerging from the rotating star hits the inner surface of the surrounding ejecta (see Fig.~\ref{fig:density}). The interaction is strongest along the equator and the first effect is a matter motion along the equator. This gives a negative signal. Later on when the jet forms along the z-axis the sign  changes as both the matter (that moves dominantly along the z axis) and the EM contribution (that moves along the z axis with a toroidal B component) give rise to a positive signal. While it is not clear that this is a generic feature, as it depends on the way that Poynting flux is injected into the system and on the configuration of the surrounding ejecta it is a remarkable demonstration of the ability of this GW signal to reveal hidden features within GRB central engines, as well as in other jetted systems.

Finally,  the {amplitude} spectral density, {$2 \sqrt{f}\,|\tilde h(f)|$ (in units of ${\rm Hz}^{-1/2}$), where $\tilde h(f)$ is the Fourier transform of $h_+(t')$,}   (see Fig.~\ref{fig:asd}) behaves at low frequencies like $f^{-1/2}$ corresponding to {$|\tilde h(f)| \propto 1/f$} - the Fourier transform of a Heaviside function. 
This reflects the memory-like nature of this signal. We note, however, that had we waited long enough, the jet would have slowed down due to interaction with surrounding matter (the GRB afterglow) and eventually the signal would vanish and so will {$\tilde h(f)$} at extremely low frequencies.

The spectral density can be converted into the energy carried by the gravitational waves. The energy flux of a GW is $\diff E/(\diff A\,\diff t) = (c^3/16\pi G)\,\langle \dot h_+^2 + \dot h_\times^2\rangle$ \citep{MTW2017}.
Integrating over the duration of the signal and using Parseval's theorem, the energy radiated per unit solid angle and per logarithmic frequency interval, at a distance $D$ from the source, is
\begin{equation}
\frac{\diff E_{\rm GW}}{\diff \Omega\, \diff \ln f} = \frac{\pi c^3}{2G}\, D^2\, f^3\, |\tilde h_+(f)|^2 ,
\label{eq:dEdlnf}
\end{equation}
where we have kept only the plus polarization computed here; the cross polarization can at most double this value. Note that the dependence on $D$ cancels, as $|\tilde h_+| \propto D^{-1}$. In terms of the amplitude spectral density shown in Fig.~\ref{fig:asd}, $2\sqrt{f}\,|\tilde h(f)|$, the right hand side is $(\pi c^3 D^2/8G)\, f^2\, [2\sqrt{f}\,|\tilde h(f)|]^2$.

\begin{figure}
  \centering
  \includegraphics[width=\columnwidth]{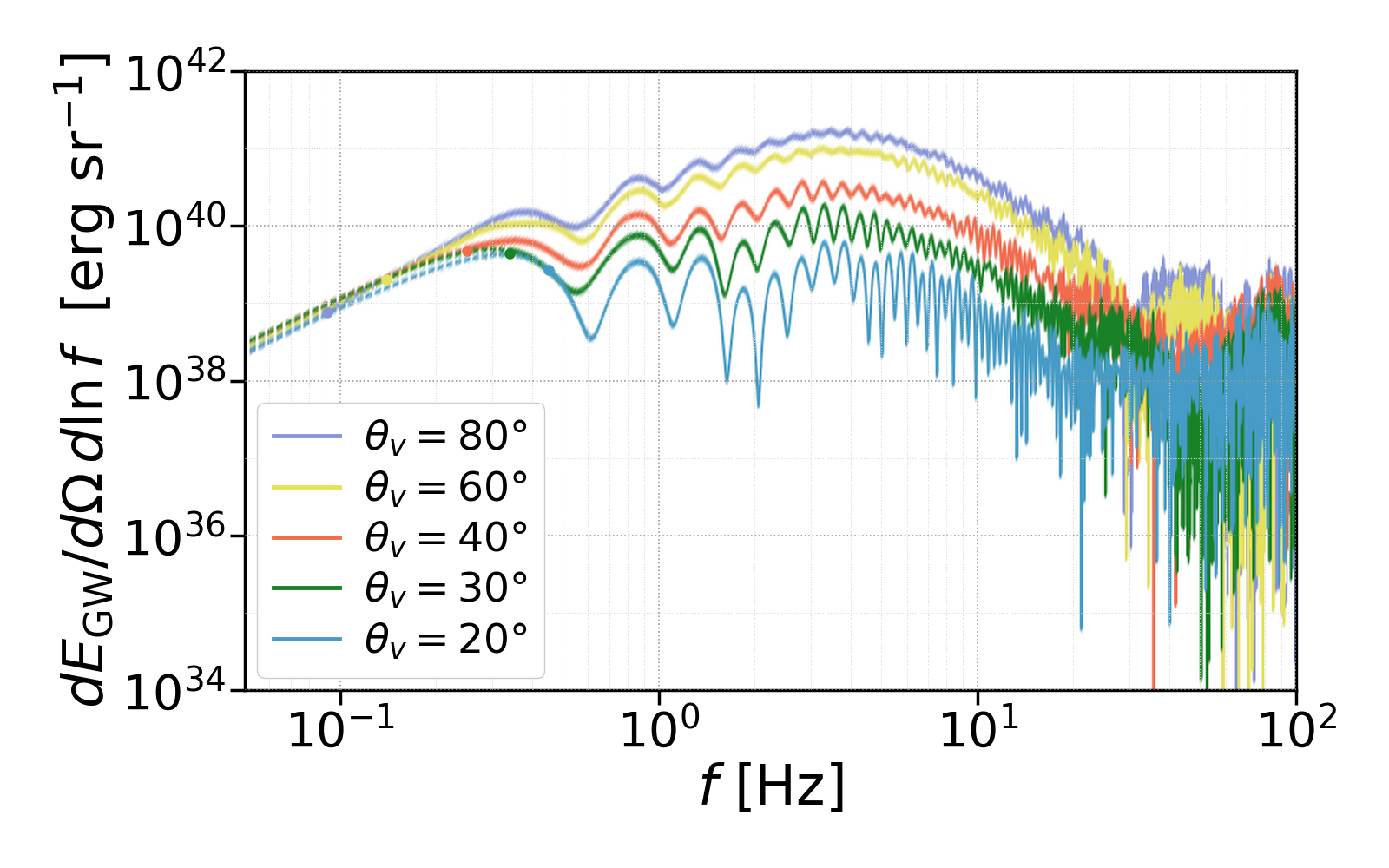}
  \caption{The gravitational-wave energy radiated per unit solid angle and per logarithmic frequency interval, $dE_{\rm GW}/d\Omega\, d\ln f$, obtained from the spectral density of Fig.~\ref{fig:asd} through Eq.~\eqref{eq:dEdlnf} for the jet of Fig.~\ref{fig:waveform}. The dots mark the truncation frequencies $1/t_{\rm cut}$ and the dashed parts the constant extrapolation of the signal, as in Fig.~\ref{fig:asd}. The spectrum peaks at $f \approx 3$--$5$~Hz, corresponding to the early dip in the waveform, and reaches $\approx 2\times 10^{41}$~erg~sr$^{-1}$ at $\theta_v = 80^\circ$. Above $\sim 20$~Hz the curves are dominated by numerical noise.}
  \label{fig:energy}
\end{figure}

Figure~\ref{fig:energy} depicts $\diff E_{\rm GW}/\diff \Omega\, \diff \ln f$ for the five viewing angles of Fig.~\ref{fig:asd}. The factor $f^3$ removes the $f^{-2}$ memory tail of $|\tilde h|^2$, and the energy spectrum rises linearly with $f$ below the knee at $\sim 0.3$--$0.5$~Hz. It peaks at $f \approx 3$--$5$~Hz, where it reaches $\approx 2\times 10^{41}$~erg~sr$^{-1}$ for $\theta_v = 80^\circ$, $\approx 10^{41}$~erg~sr$^{-1}$ for $\theta_v = 60^\circ$ and a few times $10^{40}$~erg~sr$^{-1}$ for $\theta_v = 20^\circ$--$40^\circ$. Above $\sim 20$~Hz the curves are dominated by the numerical noise of the Fourier transform, amplified here by the factor $f^3$, and carry no physical information. Integrating over the peak, the energy radiated per unit solid angle is a few times $10^{41}$~erg~sr$^{-1}$ at large viewing angles, and the total energy emitted in gravitational waves is of order $10^{42}$~erg, i.e. a fraction $\sim 10^{-9}$ of the $8\times 10^{50}$~erg injected into the jet.

Table~\ref{tab:energy} lists the energy per unit solid angle integrated over the resolved band, from the truncation frequency $f_{\rm min} = 1/t_{\rm cut}$ of each viewing angle up to the Nyquist frequency of the simulation output, $f_{\rm max} = 125$~Hz. It increases steeply with the viewing angle, approximately as $\sin^4\theta_v$, as expected for a strain amplitude $\propto \sin^2\theta_v$. The residual trend in the last column reflects the relativistic beaming factor $(1-\beta\cos\theta_v)^{-1}$, which favours small angles, and the different lower cutoffs; the $10^\circ$ entry, cut at $2$~Hz, is only a lower limit. Integrating over solid angle, counting both the jet and the counter jet, gives a total $E_{\rm GW} \approx 1.0\times 10^{42}$~erg.

\begin{deluxetable}{cccc}
\tablecaption{Gravitational-wave energy per unit solid angle emitted within the resolved frequency band, $\int_{f_{\rm min}}^{f_{\rm max}} (\diff E_{\rm GW}/\diff \Omega\, \diff \ln f)\, \diff \ln f$, for the viewing angles of Fig.~\ref{fig:energy}. The lower limit  $1/t_{\rm cut}$ is the truncation frequency marked by the dots in Fig.~\ref{fig:asd}; the upper limit is the Nyquist frequency $f_{\rm max} = 125$~Hz. \label{tab:energy}}
\tablehead{\colhead{$\theta_v$} & \colhead{$f_{\rm min}$ [Hz]} & \colhead{$dE_{\rm GW}/d\Omega$ [erg sr$^{-1}$]} & \colhead{$(dE_{\rm GW}/d\Omega)/\sin^4\theta_v$}}
\startdata
$10^\circ$ & 2      & $2.6\times10^{38}$ & $2.9\times10^{41}$ \\
$20^\circ$ & 0.455  & $3.5\times10^{39}$ & $2.5\times10^{41}$ \\
$30^\circ$ & 0.345  & $1.2\times10^{40}$ & $1.9\times10^{41}$ \\
$40^\circ$ & 0.25   & $3.1\times10^{40}$ & $1.8\times10^{41}$ \\
$60^\circ$ & 0.145  & $8.8\times10^{40}$ & $1.6\times10^{41}$ \\
$80^\circ$ & 0.0909 & $1.4\times10^{41}$ & $1.4\times10^{41}$ \\
\enddata
\end{deluxetable}

The energy is dominated by the fastest structure in the waveform rather than by the overall rise. For a memory signal of amplitude $h_0$ that changes over a time $\tau$, Eq.~\eqref{eq:dEdlnf} gives $\diff E_{\rm GW}/\diff \Omega \approx (c^3/16\pi G)\, D^2 h_0^2/\tau$. With $h_0 \approx 2G{\cal E}/(c^4 D)$ this is
\begin{equation}
\frac{\diff E_{\rm GW}}{\diff \Omega} \approx \frac{G {\cal E}^2}{4\pi c^5\, \tau} ,
\label{eq:Escaling}
\end{equation}
since the angular factor $\sin^2\theta_v/(1-\beta\cos\theta_v)$ never exceeds 2, essentially independent of the Lorentz factor. For ${\cal E} = 8\times 10^{50}$~erg and $\tau = t_{\rm inj} = 2$~s Eq.~\eqref{eq:Escaling} gives $2\times 10^{40}$~erg~sr$^{-1}$, while for the early dip of Fig.~\ref{fig:waveform}, whose amplitude is comparable to the plateau but which develops over $\tau \approx 0.04$~s, it gives $\sim 10^{42}$~erg~sr$^{-1}$. The peak of Fig.~\ref{fig:energy} at a few Hz lies between these two values, confirming that most of the GW energy is emitted during the brief interaction of the Poynting flux with the inner ejecta rather than during the 2~s injection phase.

The scaling $E_{\rm GW} \propto {\cal E}^2/\tau$ also explains why the energy found here is much smaller than the estimates of $10^{45}$--$10^{47}$~erg quoted in early studies \citep{segalis2001,piran2002}. Those estimates assume that the entire jet energy is accelerated on a millisecond timescale, a thousand times shorter than the fastest feature resolved in the present simulation, whose output cadence of 4~ms excludes any structure above $\sim 100$~Hz. If the observed millisecond variability of short GRBs originates at the central engine and is imprinted on the jet's energy flux, the emitted GW energy, and in particular its high-frequency part relevant for ground-based detectors, would be correspondingly larger. The present simulation, with its smooth injection, should therefore be regarded as a lower limit on the GW energy.

\begin{figure}
    \centering
    \includegraphics[width=1.0\linewidth]{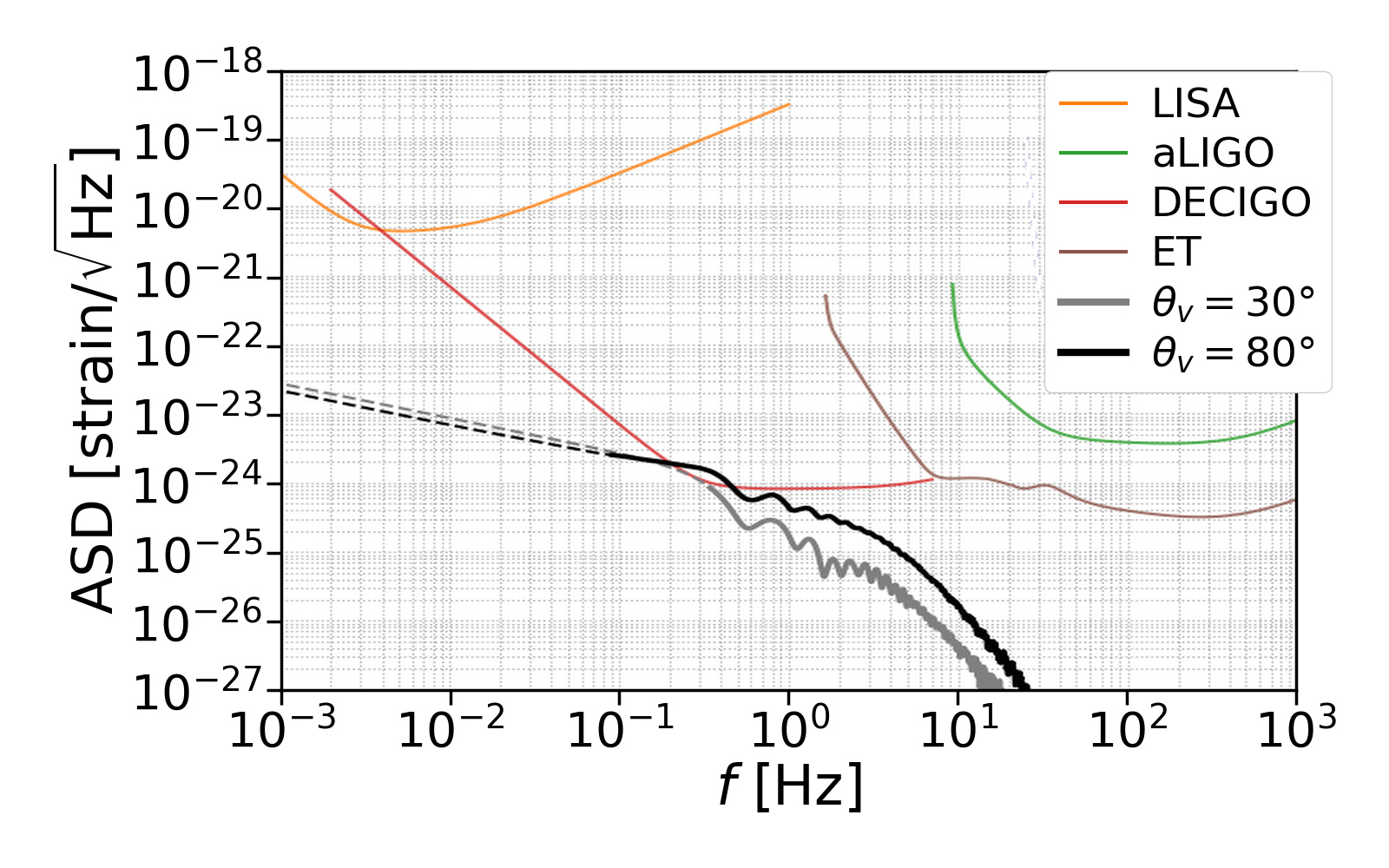}
    \caption{Amplitude spectral density (ASD) of the resulting gravitational-wave
  signal {from the jet of Fig.~\ref{fig:waveform} at $D=10$ Mpc,} for two different observer angles: $\theta_v = {80}^{\circ}$ (black),
  $\theta_v = 30^{\circ}$ (gray), shown alongside the sensitivity curves of
  various gravitational-wave detectors. The dashed parts of the GW curves mark frequencies below $1/t_{\rm cut}$, where $t_{\rm cut}$ is the time at which the signal is truncated. 
  At these frequencies the spectrum is extrapolated by the constant continuation of the signal. Similar events are only marginally
  detectable by the planned DECIGO detector{; the signal is somewhat stronger at the smaller viewing angle}.
  Detection would likely require either a significantly nearer source or a
  substantially more energetic jet.}
    \label{fig:detectability}
\end{figure}

Fig. \ref{fig:detectability} depicts a comparison of the expected spectral density with the sensitivity of several current and expected detectors. 

\section{Conclusions}
\label{sec:conclusions}
In this work we analyzed the gravitational-wave signal extracted from a
realistic MHD jet simulation.  We investigated both the waveform  and
the physical processes responsible for its evolution.

The decomposition of the stress-energy tensor into baryonic and electromagnetic
components shows that the two are comparable. As the baryonic source  is dominated by matter moving along the z-axis and the EM is dominated by a Poynting flux propagating along the z-axis with a toroidal magnetic field both contributions have the same (chosen to be positive) sign.

A pronounced global minimum appears at early observer times in the baryonic waveform. Comparison with the jet evolution suggests that this feature is associated with the transition of the magnetic flow at the onset of the simulation and with the kick that is given to the matter along the equatorial plane at this stage. This is a demonstration of the ability of the jet-GW to reveal hidden features of GRB engines.

The predicted gravitational-wave signal has a knee at {$\sim 0.3$--$0.5$ Hz, of order $1/t_{\rm inj}$}. This is determined by the duration of the energy injection chosen to be {2 s} in this example. The high-frequency tail extends up to about 10 Hz. On the lower frequency side there is a long tail with {$|\tilde h(f)| \propto 1/f$} as expected from a memory type signal that behaves like a Heaviside function. 

The total energy emitted in gravitational waves is small. Integrating the energy spectrum over frequency and solid angle (Fig.~\ref{fig:energy} and Table~\ref{tab:energy}) gives $E_{\rm GW} \approx 10^{42}$~erg, about $10^{-9}$ of the energy injected into the jet, with an angular distribution that follows $\sin^4\theta_v$. Most of this energy is emitted at a few Hz, during the brief ($\sim 0.04$~s) interaction of the Poynting flux with the inner ejecta, and not during the 2~s injection phase. This arises from  the scaling $E_{\rm GW} \propto {\cal E}^2/\tau$ for a memory-type signal, where $\tau$ is the shortest timescale over which the energy flux changes. The early estimates of $10^{45}$--$10^{47}$~erg \citep{segalis2001,piran2002} assumed that the whole jet is accelerated on a millisecond timescale. The smooth injection used here and the 4~ms output cadence do not resolve such timescales, so the present value is a lower limit. Millisecond variability at the central engine would increase both the total energy and its high-frequency part, which is the one relevant for ground-based detectors.

When comparing the spectral density with 
sensitivity curves of present and future detectors DECIGO, which is planned to operate in the deciHz range is the best detector{: an event like the one simulated here is marginally detectable by DECIGO from $\sim 10$ Mpc}. A signal with this energy of a few times $10^{50}$ erg must be at a distance of $\sim 1$ Mpc to be detected by higher-frequency detectors such as the Einstein Telescope. 
But note however, that typical short GRB signals have variable structure on a short time scale. If this variability originate at the central engine  we would have seen a higher frequency signal that was not seen here, increasing the detectability by the higher frequency detectors.  
As the signal increases linearly with the source's energy an event ten times more energetic can be detected from 10 Mpc. On the low frequency side the signal is several orders of magnitude below the sensitivity of LISA. 

While this latter result is somewhat pessimistic, it is encouraging to find that as expected the details of the waveform can reveal information on the nature and operation of hidden parts of GRB (and possibly other) jet engines. A remarkable candidate would be a hidden supernova jet \citep{Piran2019,piran2022}. Such  jets are more powerful and as the rate of these event is larger they can be expected to be found nearer than the short GRB jets discussed here. 

The research was supported by an ERC advanced grant MultiJets  and the SCEECS collaboration of Simons foundation (OZ,DDO,EE and TP) and  by an ISF grant 2067/22, a BSF grant 2024297 and NSF-BSF grant 2024788 (OB).

\bibliographystyle{aasjournal}

\begin{thebibliography}{}
\expandafter\ifx\csname natexlab\endcsname\relax\def\natexlab#1{#1}\fi
\providecommand{\url}[1]{\href{#1}{#1}}
\providecommand{\dodoi}[1]{doi:~\href{http://doi.org/#1}{\nolinkurl{#1}}}
\providecommand{\doeprint}[1]{\href{http://ascl.net/#1}{\nolinkurl{http://ascl.net/#1}}}
\providecommand{\doarXiv}[1]{\href{https://arxiv.org/abs/#1}{\nolinkurl{https://arxiv.org/abs/#1}}}

\bibitem[{Birnholtz \& Piran(2013)}]{birnholtz2013}
Birnholtz, O., \& Piran, T. 2013, Physical Review D, 87, 123007, \dodoi{10.1103/PhysRevD.87.123007}

\bibitem[{Bromberg {et~al.}(2018)Bromberg, Tchekhovskoy, Gottlieb, Nakar, \& Piran}]{bromberg2018}
Bromberg, O., Tchekhovskoy, A., Gottlieb, O., Nakar, E., \& Piran, T. 2018, Monthly Notices of the Royal Astronomical Society, 475, 2971, \dodoi{10.1093/mnras/stx3316}

\bibitem[{{Eichler} {et~al.}(1989){Eichler}, {Livio}, {Piran}, \& {Schramm}}]{Eichler1989}
{Eichler}, D., {Livio}, M., {Piran}, T., \& {Schramm}, D.~N. 1989, \nat, 340, 126, \dodoi{10.1038/340126a0}

\bibitem[{{Hotokezaka} {et~al.}(2013){Hotokezaka}, {Kiuchi}, {Kyutoku}, {Okawa}, {Sekiguchi}, {Shibata}, \& {Taniguchi}}]{Hotokezaka2013}
{Hotokezaka}, K., {Kiuchi}, K., {Kyutoku}, K., {et~al.} 2013, Physical Review D, 87, 024001, \dodoi{10.1103/PhysRevD.87.024001}

\bibitem[{Leiderschneider \& Piran(2021)}]{leiderschneider2021}
Leiderschneider, E., \& Piran, T. 2021, Physical Review D, 104, 104002, \dodoi{10.1103/PhysRevD.104.104002}

\bibitem[{{Misner} {et~al.}(2017){Misner}, {Thorne}, \& {Wheeler}}]{MTW2017}
{Misner}, C.~W., {Thorne}, K.~S., \& {Wheeler}, J.~A. 2017, {Gravitation} (Princeton University Press)

\bibitem[{Piran(2002)}]{piran2002}
Piran, T. 2002, in General Relativity and Gravitation, ed. N.~T. Bishop \& S.~D. Maharaj, 259--275, \dodoi{10.1142/9789812776556_0013}

\bibitem[{Piran(2022)}]{piran2022}
Piran, T. 2022, Jet Gravitational Waves, arXiv:2210.02740.
\newblock \doarXiv{2210.02740}

\bibitem[{{Piran} {et~al.}(2019){Piran}, {Nakar}, {Mazzali}, \& {Pian}}]{Piran2019}
{Piran}, T., {Nakar}, E., {Mazzali}, P., \& {Pian}, E. 2019, \apjl, 871, L25, \dodoi{10.3847/2041-8213/aaffce}

\bibitem[{Sago {et~al.}(2004)Sago, Ioka, Nakamura, \& Yamazaki}]{sago2004}
Sago, N., Ioka, K., Nakamura, T., \& Yamazaki, R. 2004, Physical Review D, 70, 104012, \dodoi{10.1103/PhysRevD.70.104012}

\bibitem[Sakai et al.(2025)]{Sakai2025} Sakai, Y., Yamazaki, R., Okutani, Y., et al.\ 2025, \prd, 112, 6, 064029. doi:10.1103/58pt-ms85

\bibitem[{Segalis \& Ori(2001)}]{segalis2001}
Segalis, E.~B., \& Ori, A. 2001, Physical Review D, 64, 064018, \dodoi{10.1103/PhysRevD.64.064018}

\bibitem[{{Urrutia} {et~al.}(2026){Urrutia}, {De Colle}, {Janiuk}, {Moreno}, \& {Ramirez-Ruiz}}]{Urrutia+2026_Hidden}
{Urrutia}, G., {De Colle}, F., {Janiuk}, A., {Moreno}, C., \& {Ramirez-Ruiz}, E. 2026, arXiv e-prints, arXiv:2609.17776, \dodoi{10.48550/arXiv.2609.17776}

\bibitem[{{Urrutia} {et~al.}(2023){Urrutia}, {De Colle}, {Moreno}, \& {Zanolin}}]{Urrutia+MNRAS2023}
{Urrutia}, G., {De Colle}, F., {Moreno}, C., \& {Zanolin}, M. 2023, \mnras, 518, 5242, \dodoi{10.1093/mnras/stac3433}

\bibitem[{Weinberg(1972)}]{WeinbergBook1972}
Weinberg, S. 1972, Gravitation and Cosmology: Principles and Applications of the General Theory of Relativity (New York: Wiley)

\bibitem[{Zuriel(2026)}]{zuriel2026}
Zuriel, O. 2026, {M.Sc.} thesis, The Hebrew University of Jerusalem

\end{thebibliography}

\end{document}